\documentclass{article}%
\pdfoutput=1
\usepackage{amssymb}
\usepackage{amsmath}
\usepackage{amsfonts}
\usepackage{graphicx}%
\begin{document}

\title{Induced betweenness in order-theoretic trees}
\author{Bruno Courcelle\\LaBRI, CNRS, Bordeaux University\\courcell@labri.fr}
\maketitle

\begin{quote}
\textbf{Abstract }: \emph{Betweenness} is an abstract topological notion that
has been studied for a long time in different structures.\ Informally, the
ternary relation $B(x,y,z)$ states that an element $y$ is \emph{between} $x$
and $z$, in a sense that depends on the considered structure.\ In a partially
ordered set $(N,\leq)$, $B(x,y,z):\Longleftrightarrow x<y<z\vee z<y<x$.\ The
corresponding \emph{betweenness structure} is $(N,B)$.\ The class of
betweenness structures of linear orders is \emph{first-order definable}; in
other words, it is axiomatized by a first-order sentence.\ That of partial
orders is \emph{monadic second-order} definable.\ 

An\ \emph{order-theoretic tree} is a partial order $(N,\leq)$ such that, the
set of elements larger that any element is linearly ordered, and any two
elements have an upper-bound.\ A rooted tree $T$ ordered by the ancestor
relation is an order-theoretic tree. In an order-theoretic tree, we define
$B(x,y,z)$ to mean that $x<y<z$\ or $z<y<x$\ or $x<y\leq x\sqcup z$\ or
$z<y\leq x\sqcup z$\ provided the least upper-bound $x\sqcup z$ of $x$ and $z$
is defined when $x$ and $z$ are incomparable. In a previous article, we
established that the corresponding class \textbf{BO}\ of betweenness
structures is monadic second-order definable. We left as a conjecture that the
class \textbf{IBO} of \emph{induced substructures }of the structures in
\textbf{BO} is monadic second-order definable. We prove this conjecture.\ Our
proof uses \emph{partitioned probe cographs} (related to cographs), and their
six finite minimal excluded induced subgraphs called their \emph{bounds}%
.\ This proof links two apparently unrelated topics: cographs and
order-theoretic trees.

However, the class \textbf{IBO}\ has finitely many \emph{bounds},\emph{ i.e.},
minimal excluded finite induced substructures.\ Hence it is first-order
definable.\ The proof of finiteness uses well-quasi-orders and does not
provide the finite list of bounds.\ Hence, the associated first-order defining
sentence is not yet known.\ 
\end{quote}

{\LARGE Introduction}

Betweenness is an abstract topological notion that has been studied for a long
time in different structures\ \cite{Cha+,Chv,Hu,Pi,Sho1,Sho2}. Informally, the
ternary relation of \emph{betweenness} $B(x,y,z)$ states that an element $y$
is \emph{between} $x$ and $z$ in a sense that depends on the considered
structure.\ In a partially ordered set $(N,\leq)$,
$B(x,y,z):\Longleftrightarrow x<y<z\vee z<y<x$.\ The corresponding
\emph{betweenness structure} is $(N,B)$.\ The class of betweenness structures
of linear orders is \emph{first-order definable}, \emph{i.e.}, is axiomatized
by a single first-order sentence; that of partial orders is \emph{monadic
second-order} definable \cite{CouBetPO,CouEng,Lih}.\ 

The betweenness structures of certain finite graphs have been studied in
\cite{Cha+,Chv} and those of trees of various kinds
in\ \cite{Cou14,CouYuri,CouLMCS2020}.

An \emph{order-theoretic forest} (an \emph{O-forest} in short) is a partial
order $T=(N,\leq)$ such that $\{x\in N\mid x\geq y\}$ is linearly ordered for
any $y$, a notion studied by Fra\"{\i}ss\'{e} \cite{Fra} (under the name of
\emph{tree}). It is an \emph{order-theoretic tree} (an \emph{O-tree} in short)
if any two elements have an upper-bound.\ A rooted tree ordered by the
ancestor relation is an O-tree.\ The ordered set of rational numbers is an
O-tree in which no node has an immediate ancestor. In previous articles, we
used O-trees to define the modular decomposition and the rank-width of
countable graphs \cite{Cou14,CouDel}.\ 

The \emph{betweenness structure} of an O-forest $T=(N,\leq)$ is $(N,B_{T}\ )$
such that

\begin{quote}
$B_{T}(x,y,z):\Longleftrightarrow x\neq y\neq z\neq x\wedge\lbrack(x<y\leq
x\sqcup z)\vee(z<y\leq x\sqcup z)]$
\end{quote}

where $x\sqcup z$ denotes the least upper-bound of $x$ and $z$. If $x\sqcup z$
does not exist, there is no triple $(x,y,z)$ in $B_{T}$. We denote by
\textbf{BO}\ this class of structures.\ An \emph{induced betweenness }is a
induced substructure of such $(N,B_{T})$ where $T$ can be, equivalently, an O-tree.

In \cite{Cou14,CouLMCS2020}, we have axiomatized by \emph{monadic second-order
sentences} several classes of betweenness structures related to O-trees, in
particular \textbf{BO}. We conjectured that the class \textbf{IBO} of
\emph{induced} betweenness structures in O-trees is monadic second-order
definable too. In the present article, we prove a slight weakening of this
conjecture by allowing the defining monadic second-order sentence to use the
finiteness set predicate $Fin(X)$, expressing that a set $X$ is finite.
Furthermore, as we explain below, this class is \emph{first-order definable,}
however, the proof does not construct the defining sentence.

The proof of monadic second-order definability uses \emph{partitioned probe
cographs}.\ They are related as follows to \emph{cographs}, the graphs without
induced paths $P_{4}$.\ A \emph{probe cograph} is obtained from a cograph by
choosing a subset $L$ of its vertex set and by removing the edges whose two
ends are in $L$.\ In a partitioned probe cograph, we keep track of $L$\ by
labelling its vertices. A probe cograph is obtained from a partitioned probe
cograph by forgetting such labels. The path $P_{5}$ is a probe cograph.\ The
path $P_{6}$\ is not.

The class of partitioned probe cographs is \emph{hereditary}, \emph{i.e.}, is
closed under taking induced subgraphs.\ A \emph{bound} of such a class
$\mathcal{C}$\ is a minimal induced subgraph that is not in $\mathcal{C}$, a
terminology used by Pouzet \cite{Pou}. Partitioned probe cographs have six
bounds, determined in \cite{VBL}.\ 

The class of probe cographs is also hereditary.\ As it is
2-\emph{well-quasi-ordered} \cite{Dal+,Pou}, it has finitely many bounds,
among which the path $P_{6}$. We exhibit a few others, without proving that
the list is complete. This list is not completely known.\ We define an
algorithm, based on monadic second-order logic and decompositions of the
graphs as tree-like unions of complete bipartite graphs that \emph{could}
compute them. Unfortunately, this algorithm is intractable.

Our proof links two apparently unrelated notions: betweenness in O-trees and a
variant of cographs.\ To give an intuition about this link, we observe that
the cocomparability graph of an O-tree is a graph without induced $P_{4}$,
hence, is a possibly infinite cograph (Proposition 3.5).\ 

Finally, we use a result by Pouzet \cite{Pou} to prove that the class
\textbf{IBO}\ of ternary structures has finitely many \emph{bounds},\emph{
i.e.,} minimal excluded induced finite substructures, hence is first-order
definable.\ This proof uses well-quasi-orders of labelled ternary structures,
and does not provide the finite list of bounds.\ Hence, the associated
first-order defining sentence is not known.\ We have no algorithm, even
intractable (as for probe cographs), to determine the bounds of \textbf{IBO}%
.\ We explain this fact by exhibiting a property of clique-width that does not
extend from graphs to ternary structures, as we would need.

To summarize, we consider the following hereditary classes of graphs or
ternary structures having finitely many bounds:

cographs, the unique bound is $P_{4}$;

partitioned probe cographs, they have six known bounds;

probe cographs, the finite set is unknown but computable;

\textbf{IBO}, the finite set is unknown and we have no algorithm, even
intractable, to compute it.

\bigskip

\textbf{Summary:} Section 1 reviews partial orders, graphs, clique-width,
trees and logic.\ Section 2\ defines and studies probe cographs.\ Section 3
defines order-theoretic trees.\ Section 4 proves the two logical
characterizations of \textbf{IBO}.\ Section 5 exhibits some bounds of the
class of probe cographs. Section 6 states and discusses three open problems.

\bigskip

\textbf{Acknowledgement:} I thank Maurice Pouzet for fruitful email exchanges,
from which I obtained the FO-definability of the class \textbf{IBO}, by a
proof based on one of his works. I thank the referees for many useful
comments, especially regarding references.

\section{Definitions and notation}

In this article, all graphs, trees, partial orders and, more generally, all
relational structures are \emph{countable}, which means finite or countably
infinite\footnote{The restriction to countable structures makes it possible to
define effective descriptions and to obtain computability results.\ For
example, it is decidable whether two regular join-trees are isomorphic
\cite{CouLMCS}, Corollary 3.31.}. We denote by $[k]$ the set $\{1,...,k\}$%
.\ In some cases, we denote by $A\uplus B$\ the union of two sets $A$ and $B$
to stress that they are disjoint.\ 

\bigskip

\textbf{Partial orders}

For a partial order $\leq,\subseteq$, we denote respectively by $<,\subset$,
the corresponding strict partial order. We write $x\bot y$ if $x$ and $y$ are
incomparable for the considered order.

Let $P=(V,\leq)$ be a partial order.\ For $X,Y\subseteq V$, the notation $X<Y$
means that $x<y$ for every $x\in X$ and $y\in Y$.\ We write $X<y$ instead of
$X<\{y\}$ and similarly for $x<Y$. We use a similar notation for $\leq$\ and
$\bot$. The least upper-bound of $x$ and $y$ is denoted by $x\sqcup y$ if it
exists and is then called their \emph{join}.\ 

If $X\subseteq V$, then we define $N_{\leq}(X):=\{y\in V\mid y\leq X\}$\ and
similarly for $N_{<}$. We have $N_{\leq}(X)\leq X$ and $N_{\leq}(\emptyset
)=V$. We define $L_{\geq}(X):=\{y\in V\mid y\geq X\}$, and similarly
$L_{>}(X).$ We write $L_{\geq}(x)$ (resp. $L_{\geq}(x,y))$ if $X=\{x\}$ (resp.
$X=\{x,y\}$) and similarly for $L_{>}.$

An \emph{embedding} of a partial order $P=(N,\leq)$ into another one
$P^{\prime}=(N^{\prime},\leq^{\prime})$ is an injective mapping
$h:N\rightarrow N^{\prime}$ such that $h(x)\leq^{\prime}h(y)$ if and only if
$x\leq y$. Hence it is monotone (or isotone).\ It is a \emph{join-embedding}
if $h(x)\sqcup^{\prime}h(y)=h(x\sqcup y)$ whenever $x$ and $y$ have a join in
$P$. We write $P\subseteq_{j}P^{\prime}$ if $N\subseteq N^{\prime}$ and the
inclusion mapping is a join-embedding. If $P$ and $P^{\prime}$ are labelled,
then the labels are preserved in embeddings.

\bigskip

\textbf{Graphs}

Graphs are undirected and simple, which means without loops and parallel
(multiple) edges. We denote respectively by $P_{n},C_{n},K_{n}$ a path, a
cycle and a clique with $n$ vertices.

The notation $u-v$ designates an edge with ends $u$ and $v$.\ As a property,
it means also "there is an edge between $u$ and $v$". We say then that $u$ and
$v$ are \emph{adjacent} or \emph{are neighbours}.\ The notation $u-v-w-x\ $%
shows the vertices and edges of a path of 4 vertices. The notation $u-v-w-x-u$
shows the vertices and edges of a 4-cycle.

Induced subgraph inclusion is denoted by $\subseteq_{i}$ and $G[X]$ is the
induced subgraph of $G=(V,E)$ with vertex set $X$ $\subseteq V$.\ Then,
$G-x:=G[V-\{x\}]$.

We denote by $G\oplus H$\ the union of disjoint graphs $G$ and $H$. We define
$G\otimes H$ as $G\oplus H$\ augmented with edges between any vertex of $G$
and any vertex of $H$. We can use the notation $G_{1}\oplus G_{2}%
\oplus...\oplus G_{n}$ because the operation $\oplus$ is associative, and
similarly for $\otimes.$ We can also use the notation $\oplus(G_{1}%
,G_{2},...,V_{n})$ or $\otimes(G_{1},G_{2},...,G_{n})$ if we consider $t$ as a
rooted tree whose internal vertices are labelled by $\oplus$\ or $\otimes$.

The \emph{diameter} of a connected graph is the maximal \emph{distance}
between two vertices, \emph{i.e.}, the minimum number of edges of a path
between them.

By a \emph{class} of graphs, we mean a set closed under isomorphism.\ 

\bigskip

\textbf{Rooted trees}

In Graph Theory, a \emph{tree} is a connected graph without cycles. It is
convenient to call \emph{nodes} the vertices of a tree because in some proofs,
we will discuss simultaneously a graph and an associated tree. A \emph{rooted
tree} is a triple $T=(N_{T},E_{T},r)$ such that $(N_{T},E_{T})$ is a tree and
$r\in N$ is a distinguished node called its \emph{root}. This tree can be
defined from the partial order $(N_{T},\leq_{T})$ such that $x\leq_{T}y$ if
and only if $y$ is on the path from the \emph{root} $r$ to $x$. In most cases,
we will handle a rooted tree $T$\ as a partial order $(N_{T},\leq_{T})$.\ In
Section 3, we will generalize rooted trees into \emph{order-theoretical}
\emph{trees}, defined as partial orders, as done by Fra\"{\i}ss\'{e}
\cite{Fra}.

A \emph{leaf} is a minimal node and $L_{T}$ denotes the set of leaves.\ The
other nodes are \emph{internal}.

If $x<_{T}y$, then $y$ is an \emph{ancestor} of $x$.\ \ A node $x$ is a
\emph{son} of $y$ if $x<_{T}y$ and there is no node $z$ such that
$x<_{T}z<_{T}y$. The \emph{degree} of a node is the number of its sons.\ A
node of degree 0 is thus a leaf.

The \emph{subtree} of a rooted tree $T=(N,\leq)$ \emph{issued }from a node $u$
is $T/u:=(N_{\leq}(u),\leq^{\prime})$ where $\leq^{\prime}$ is the restriction
of $\leq$ to $N_{\leq}(u).$

A \emph{rooted forest} $F$\ is the union of pairwise disjoint rooted trees and
$Rt_{F}$ denotes the set of roots of its trees.

A finite rooted tree $T$ can be denoted linearly by $\widetilde{T}$ defined as
follows (which is useful in inductive proofs) :\ 

\begin{quote}
if $T$ is reduced to $r$, then $\widetilde{T}:=r,$

if $T$ has root $r$ and subtrees $T_{1},...,T_{p}$ issued from the sons of the
root, then $\widetilde{T}:=r(\widetilde{T_{1}},...,\widetilde{T_{p}})$.\ 
\end{quote}

Any permutation of the sequence $T_{1},...,T_{p}$ defines the same tree
because there is no defined order between the sons of a node.

\bigskip

\textbf{Relational structures and logic}

A \emph{relational structure }is a tuple $S=(D,R_{1},...,R_{p})$ where $D$ is
a set, its \emph{domain} and $R_{1},...,R_{p}$ are relations of fixed
arity.\ The \emph{signature} of $S$ is the sequence of arities of the
relations $R_{1},...,R_{p}.$ We will consider classes of structures having a
fixed signature.

If $S$ is a relational structure with domain $N$\ and $X\subseteq N$, then
$S[X]$ denotes the induced substructure with domain $X,$ and $\subseteq_{i}$
denotes an induced inclusion of relational structures of same signature.

We will use structures $(N,\leq)$ to describe a partial order, a rooted tree
or an order-theoretic forest (defined in Section 3), $(V,edg)$ to describe an
undirected graph with set $V$ of vertices where $edg(x,y)$ means that there is
an edge between $x$ and $y$, and\ $(N,B)$ for a\ \emph{betweenness structure}
(cf.\ Introduction and Section 4), where $B$ is a ternary relation. Additional
unary relations will formalize labellings of the elements of $N$ or $V$.

The isomorphism of relational structures is denoted by $\simeq$. The
isomorphism class of a structure $S$ is denoted by $[S]_{\simeq}$. A set of
structures is called a \emph{class} if it is closed under isomorphism. We say
that it is \emph{finite} if the set of its isomorphism classes is finite. A
class $\mathcal{C}$ is \emph{hereditary }if it is closed under taking induced substructures.\ 

Properties of structures (and of graphs) will be expressed by
\emph{first-order (FO}) or \emph{monadic-second order (MSO)} formulas and
sentences.\ A \emph{sentence} is a formula without free variables.\ For an
example, that a graph has no induced subgraph isomorphic to a finite graph $H$
is FO expressible. That a graph is not connected is expressed in its
representing structure $(V,edg)$ by the following MSO-sentence:

$\exists X[(\exists x.x\in X)\wedge(\exists y.y\notin X)\wedge(\forall
x,y.(x\in X\wedge y\notin X\Longrightarrow\lnot edg(x,y))].$

The book \cite{CouEng} contains a detailed study of monadic second-order logic.

We will consider classes of countable (which means "finite or countably
infinite") structures.\ Such a class is \emph{MSO} (or \emph{FO})
\emph{definable} if it is the class of countable models of an MSO (or
FO)-sentence. It is \emph{uFO} \emph{definable}, if it is defined by a
\emph{universal FO-sentence}, \emph{i.e.}, of the form $\forall
x,y,z...\varphi(x,y,z,...)$ where $\varphi(x,y,z,...)$ is quantifier-free.

The finiteness of an arbitrary set\ $X$\ is not MSO expressible.\ However it
is if some linear order on $X$\ can be defined by MSO-formulas in the case
where $X$\ is part of the domain of a structure with a nonempty signature (see
Example 1.6 in \cite{CouLMCS2020}).\ This is the case for an example if $N$ is
the set of nodes of a tree of bounded degree. An MSO$_{fin}$-sentence is an
MSO-sentence where the \emph{finiteness set predicate} $Fin(X)$ expressing
that a set $X$ is finite can be used.

These definitions and the next proposition apply to graphs represented by
structures $(V,edg)$.

A class of structures $\mathcal{C}$\ is \emph{finitary }if a structure
$S$\ belongs to $\mathcal{C}$ if and only if all its \emph{finite} induced
substructures belong to $\mathcal{C}$.\ This implies that $\mathcal{C}$ is
hereditary and characterized by its subclass $\mathcal{C}_{fin}$ of finite
structures.\ If $\mathcal{C}$\ is finitary and contains an infinite structure,
then $\mathcal{C}_{fin}$ is hereditary but not finitary.

If $\mathcal{C}$ is finitary, then its \emph{bounds,} forming the class
$Bnd(\mathcal{C}),$\ are the finite structures not in $\mathcal{C}$ whose
proper induced substructures are all in $\mathcal{C}$.\ Then $\mathcal{C}$\ is
the class of structures having no induced substructure isomorphic to one in
$Bnd(\mathcal{C}).$

A class may be hereditary while having infinitely many bounds.\ Consider for
an example the class of graphs without cycles whose vertices have all degree
2; then each cycle $C_{n}$, where $n\geq3$\ is a bound of this class.

\bigskip

A routine proof can establish the following.

\bigskip

\textbf{Proposition 1.1}\ : A class of structures is uFO definable if and only
if it is finitary and has a finite set of bounds (where finiteness is up to
isomorphism). If it is so, its finite structures can be recognized in
polynomial time.

\section{Cographs and related notions}

In this section all graphs (and trees) are finite. \ 

\bigskip

\textbf{Definition 2.1\ }\ : \emph{Cographs}

(a) A graph is a \emph{cograph} if and only if it can be generated from
isolated vertices by the operations $\oplus$\ and $\otimes,$ if and only if it
has no induced path $P_{4}$. There are many other characterizations
\cite{Wiki}. The family of cographs is hereditary.\ Its only bound is $P_{4}$.

(b) Cographs can thus be defined by algebraic terms over $\oplus$\ and
$\otimes$ and nullary symbols denoting vertices. For example, the cycle
$a-b-c-d-a$ is defined by the term $(a\oplus c)\otimes(b\oplus d)$. To define
it up to isomorphism, that is, without naming the vertices, we will use the
term\ $(\bullet\oplus\bullet)\otimes(\bullet\oplus\bullet)$.

(c) We can use the notation $t_{1}\oplus t_{2}\oplus...\oplus t_{n}$ because
the operation $\oplus$ is associative, and similarly for $\otimes.$ We can
also use the notation $\oplus(t_{1},t_{2},...,t_{n})$ or $\oplus(t_{1}%
,t_{2},...,t_{n}).$

(d) The syntactic tree of a term defining a cograph $G=(V,E)$ is called a
$\{\oplus,\otimes\}$-\emph{tree}.\ It is a rooted tree whose set of leaves is
$V$ and whose internal nodes are of degree at least 2 and labelled by $\oplus
$\ or $\otimes.$

\bigskip

\textbf{Definition 2.2}~: \emph{2-graphs}

A \emph{2-graph} is a graph $(V,E)$ equipped with a bipartition $V_{1}\uplus
V_{2}$ of its vertex set $V$. We will say that $x\in V_{i}$ is an
$i$-\emph{vertex}. The \emph{type} of a finite path $x_{1}-x_{2}-...-x_{n}$ in
a 2-graph is the word $b_{1}b_{2}...b_{n}$ over $\{1,2\}$ such that $x_{j}$ is
a $b_{j}$-vertex for each $j=1,...,n$. $\ \ \ \ $

\bigskip

\textbf{Definitions 2.3 :} \emph{ Probe cographs }

(a) A \emph{partitioned probe cograph} (a \emph{pp-cograph} in short) is a
2-graph obtained from a cograph $(V,E)$ by choosing a bipartition $V_{1}\uplus
V_{2}$ of $V$ and removing the edges between its 1-vertices.\ 

(b) A \emph{probe cograph} (a \emph{p-cograph} in short) is obtained from a
pp-cograph by forgetting the bipartition (and the corresponding labelling of
its vertices by 1 or 2).

(c) A bipartition of a graph (or its corresponding vertex-labelling by 1 or 2)
is \emph{good} if it makes it into a pp-cograph.

(d) ~Partitioned probe\emph{ }cographs can be defined by terms, similar to
those that define cographs, using the operation $\oplus$ and the operation
$\otimes$ that we redefine as follows for 2-graphs: $G\otimes H$ is $G\oplus
H$ augmented with all edges between an $i$-vertex of $G$ and a $j$-vertex of
$H$, provided $i$ and $j$ are not both 1. These two operations do not modify
the vertex labellings of $G$ and $H$.\ They are associative.\ A nullary symbol
$\bullet_{i}(x)$ defines $x$ as an isolated $i$-vertex. \ \ $\square$

\bigskip

The path $P_{4}=a-b-c-d$ with labelling of type 1212 is a pp-cograph defined
by the term~$\bullet_{1}(c)\otimes([\bullet_{1}(a)\otimes\bullet_{2}%
(b)]\oplus\bullet_{2}(d))$. To define it up to isomorphism, we can use the
term\ ~$\bullet_{1}\otimes([\bullet_{1}\otimes\bullet_{2}]\oplus\bullet_{2})$.
Note that ~$\bullet_{1}(x)\otimes\bullet_{1}(y)$ and $\bullet_{1}%
(x)\oplus\bullet_{1}(y)$ define the same 2-graph. See also Example 2.5(1).
$\ \square$

\bigskip

We review some results from\ \cite{Dal+,VBL}.

\bigskip

\textbf{Proposition 2.4 : }(1) The class of partitioned probe cographs is
hereditary.\ Its bounds are the paths of types 11, 2222, 1222, 2122 or 21212
and the 2-graph $Q\ $defined as the path $a-b-c-d-e$ of type 12221 augmented
with the edge $b-d$. Partitioned probe graphs can be recognized in linear time.

(2) The class of probe cographs is hereditary and has finitely many bounds.
Its graphs can be recognized in linear time.\ 

\bigskip

An immediate consequence of interest for the present article is that
pp-cographs are uFO definable among 2-graphs.\ The defining sentence is
effectively constructed from the six \emph{known} bounds.\ Probe cographs are
so, but the corresponding uFO sentence is not known, because the complete list
of bounds is not either.\ However, their bounds are definable by a known
MSO-sentence obtained from the FO-sentence that defines the pp-cographs. We
will discuss these points in Section 5.

\bigskip

\textbf{Examples 2.5:} (1) The path $P_{4}$ is not a cograph.\ It has good
labellings of types 1212 and 1221.\ Its labellings of type 1222, 2122 and 2222
are not good.

(2) The labelled path $P_{5}=a-b-c-d-e$ of type 12121 is a pp-cograph defined
by the term $\bullet_{1}(c)\otimes\lbrack(\bullet_{1}(a)\otimes\bullet
_{2}(b))\oplus(\bullet_{1}(e)\otimes\bullet_{2}(d)]$. No other labelling of it
is good, which follows from Proposition 2.4(1).

(3) The path $P_{6}=a-b-c-d-e-f$ is not a p-cograph.\ Assume it has a good
labelling.\ The induced path $a-b-c-d-e$ must have type 12121 and $f$ must
have label 2.\ But then $b-c-d-e-f$ has type 21212, which is not possible by
(1).\ It follows that a p-cograph has no induced $P_{6}$.\ Hence, a connected
p-cograph cannot have diameter 5 or more because otherwise, it would contain
an induced path $P_{6}$. Furthermore, $P_{6}$\ is a bound of p-cographs.\ 

(4) A similar proof using (1) shows that the cycle $C_{5}$\ is a bound of
p-cographs. All other graphs having at most 5 vertices are
p-cographs.\ Proposition 5.5 will present some bounds for p-cographs.
$\ \ \ \square$

\section{Order-theoretic trees}

\textbf{Definition 3.1}: \emph{Order-theoretic forests and trees}.

(a) An \emph{order-theoretic forest} (an \emph{O-forest} in short) is a
partial order $J=(N,\leq)$ such that, for each $x\in N$, called the set of
\emph{nodes}, the set $L_{\geq}(x):=\{y\mid y\geq x\}$ is linearly
ordered.\ An O-forest is an \emph{O-tree} if every two nodes have an
upper-bound. An O-tree is a \emph{join-tree\footnote{We used join-trees to
define the modular decomposition and the rank-width of countable graphs
\cite{Cou14,CouDel}.We studied them in algebraic and logical perspectives in
\cite{CouLMCS}.}} if every two nodes $x$ and $y$ have a least upper-bound,
denoted by $x\sqcup y$ and also called their \emph{join}. An O-tree may have
no largest node. Its largest node if it exists is called the \emph{root}. If
$x\sqcup y$ and $y\sqcup z$ are defined, then so is $x\sqcup z$ and it belongs
to $\{x\sqcup y,y\sqcup z\}.$

(b) If $u<w$, then we say that $w$ is an \emph{ancestor} of $u$.

(c) A \emph{line} in an O-forest\ $(N,\leq)$ is a subset $L$ of $N$ that is
linearly ordered and \emph{convex},\emph{ i.e.}, is such that $z\in L$ if
$x,y\in L$ and $x<z<y$.

(d) A \emph{leaf} is a minimal node.\ It has \emph{degree} 0; the set of
leaves is denoted by $L_{J}$.\ 

(e) A node $x$ has \emph{degree} 1 if there is $y<x$ such that every node
$z<x$ is comparable with $y$. For finite forests, this is equivalent to the
definition in Section 1.\ If we delete some nodes of degree 1 of an
O-forest\ $J,$ we obtain a (possibly empty) O-forest $J^{\prime}$ that
join-embeds into $J$ (cf.\ Section 1) because a node of degree 1 is not the
join of any two incomparable nodes.$\ \square$

\bigskip

The partial order $(N_{T},\leq_{T})$ associated with a rooted tree $T$\ is a
join-tree such that $L_{\geq}(x)$\ is finite for each node $x$.\ Conversely,
every O-tree having this property is associated in this way with a rooted tree.

\bigskip

\textbf{Definition 3.2 }: \emph{Substitutions of lines in O-forests.}

Let $J=(N,\leq)$ be an O-forest and, for each $x\in N$, let $(A_{x},\leq_{x})$
be a (possibly empty) linearly ordered set. These sets are assumed to be
pairwise disjoint. We let $J^{\prime}=J[x\longleftarrow A_{x};x\in
N]:=(N^{\prime},\leq^{\prime})$ be the partial order such that :

\begin{quote}
$N^{\prime}$ is the union of the sets $A_{x}$,

$u\leq^{\prime}v$ if and only if either $u\leq_{x}v$ or $u\in A_{x}\wedge v\in
A_{y}\wedge x<y$, for some $x,y$.
\end{quote}

It is an O-forest in which each nonempty set $A_{x}$ is a line.

\bigskip

\textbf{Definitions 3.3\ \ }: \emph{The join-completion of an O-forest}.

Let\ $J=(N,\leq)$ be an O-forest and $\mathcal{K}$\ be the set of upwards
closed lines of the form $L_{\geq}(x,y):=L_{\geq}(x)\cap L_{\geq}(y)$\ for all
(possibly equal) nodes $x,y.\ $If $x$ and $y$ have a join, then $L_{\geq
}(x,y)=L_{\geq}(x\sqcup y)$.\ If they have no upper-bound, then $L_{\geq
}(x,y)$\ is empty.\ 

The family $\mathcal{K}$ is countable.\ We let $h:N\rightarrow\mathcal{K}%
$\ map $x$ to $L_{\geq}(x)$ and $\widehat{J}:=(\mathcal{K},\supseteq)$. We
call $\widehat{J}$ the \emph{join-completion of }$J$ because of the following
proposition, stated with these hypotheses and notation.

\bigskip

\textbf{Proposition 3.4 } \cite{CouLMCS2020} : The partially ordered set
$\widehat{J}:=(\mathcal{K},\supseteq)$ is a join-tree and $h$ is a
join-embedding $J\rightarrow\widehat{J}$. $\ \ \square$

\bigskip

If we identify $x\in N$\ with $h(x):=L_{\geq}(x),$ then $h$ defines a
join-embedding of $J$ into $\widehat{J}$. The join of $h(x)$ and $h(y)$ is
$L_{\geq}(x,y).$

\bigskip

The following side proposition shows that cographs arise naturally from
O-forests. We recall that $\bot$ denotes incomparability in a partial order.

\bigskip

\textbf{Proposition 3.5:} The cocomparability graph $CC(J):=(N,\bot)$ of a
finite forest $J=(N,\leq)$ is a cograph.

\textbf{Proof sketch:} First we prove that the cocomparability graph
$CC(T)=(N,\bot)$ of a finite rooted tree $T=(N,\leq)$ is a cograph. If
$T=a(T_{1},....,T_{n})$ and $n\geq2$, then $CC(T)=a\oplus(CC(T_{1}%
)\otimes...\otimes CC(T_{n}))$. If $n=1$, we have $CC(T)=a\oplus CC(T_{1})$.
$\ $

If $J=(N,\leq)$ is a finite forest, it is the disjoint union of rooted trees
$T_{1},...,T_{n}$, then $CC(J)=CC(T_{1})\otimes...\otimes CC(T_{n})$.
$\ \ \ \square$

\bigskip

If we define as a cograph any finite or infinite graph without induced path
$P_{4}$, then this proposition extends to countable O-forests.

\section{Betweenness in order-theoretic trees}

We will consider ternary structures $S=(N,B).$ If $n>2$, the notation
$\neq~(x_{1},x_{2},$ \ $...,x_{n})$\ \ means that $x_{1},x_{2},...,x_{n}$ are
pairwise distinct, hence it abreviates an FO-formula. If $n>3$, then
$B^{+}(x_{1},x_{2},...,x_{n})$ abreviates the FO-formula

\begin{quote}
$B(x_{1},x_{2},x_{3})\wedge B(x_{2},x_{3},x_{4})\wedge...\wedge B(x_{n-2}%
,x_{n-1},x_{n})$
\end{quote}

and $A(x_{1},x_{2},x_{3})$ abreviates

\begin{quote}
$B(x_{1},x_{2},x_{3})\vee B(x_{2},x_{1},x_{3})\vee B(x_{1},x_{3},x_{2}).$
\end{quote}

\bigskip

\textbf{Definitions and background 4.1 :} \emph{Betweenness in O-forests.}

(a) The \emph{betweenness relation} of an O-forest $J=(N,\leq)$ is the ternary
relation $B_{J}\subseteq N^{3}$ such that :

\begin{quote}
$B_{J}(x,y,z):\Longleftrightarrow\ \neq(x,y,z)\wedge([x<y\leq x\sqcup
z]\vee\lbrack z<y\leq x\sqcup z]).$
\end{quote}

We have $B_{J}(x,y,z)$ if $x<y<z.$\ If $x\sqcup z$ is undefined, then
$B_{J}(x,y,z)$ holds for no triple $(x,y,z)$.

We denote by \textbf{BO}\ the class of betweenness structures $(N,B_{J})$ of
O-forests $J=(N,\leq).$

(b) The following related classes have been considered in
\cite{CouYuri,CouLMCS2020}.

\begin{quote}
\textbf{IBO}\ is the class of induced substructures of the structures in
\textbf{BO}.

\textbf{QT}\ (for \emph{quasi-tree}s\footnote{Introduced in \cite{Cou14}.\ })
is\ the class of betweenness structures of join-trees.

\textbf{IBQT}\ is the class of induced substructures of structures in
\textbf{QT}.
\end{quote}

We have the following proper inclusions in \cite{CouLMCS2020} :

\begin{quote}
\textbf{BO} $\subset$ \textbf{IBO, IBQT}$\ \subset$ \textbf{IBO} and \textbf{
QT} $\subset$ \textbf{IBQT\ }$\cap\ $\textbf{BO}.
\end{quote}

The classes \textbf{IBQT} and \textbf{BO} are incomparable, and for finite
structures, we have \textbf{QT} $=$ \textbf{BO}.

(c) The betweenness relation $B$ of a rooted tree $T=(N,\leq)$, (hence $(N,B)$
$\in$\textbf{ QT}) satisfies the following properties for all $x,y,z,u\in N$:

\begin{quote}
A1 : $B(x,y,z)\Rightarrow$ $\neq(x,y,z).$

A2 : $B(x,y,z)\Rightarrow B(z,y,x).$

A3 : $B(x,y,z)\Rightarrow\lnot B(x,z,y).$

A4 : $B(x,y,z)\wedge B(y,z,u)\Rightarrow B^{+}(x,y,z,u).$

A5 : $B(x,y,z)\wedge B(x,u,y)\Rightarrow B^{+}(x,u,y,z).$

A6 : $B(x,y,z)\wedge B(x,u,z)\Rightarrow y=u\vee B^{+}(x,u,y,z)\vee
B^{+}(x,y,u,z).$

A7 : $\neq(x,y,z)\Rightarrow A(x,y,z)\vee\exists w~[B(x,w,y)\wedge
B(y,w,z)\wedge B(x,w,z)].$
\end{quote}

Conversely, every ternary structure satisfying these properties is in
\textbf{QT} \cite{Cou14}. Hence, the class \textbf{QT} is FO-definable.\ It is
not hereditary.\ Its closure under taking induced substructures, denoted by
\textbf{IBQT}, is uFO definable by Proposition 2.12\ of \cite{CouLMCS2020}%
.\ It is defined by A1-A6\ together with :

\begin{quote}
A8 : $\forall u,x,y,z~[\neq(u,x,y,z)\wedge B(x,y,z)\wedge\lnot
A(u,y,z)\Rightarrow B(x,y,u)]$.\ 
\end{quote}

The class \textbf{BO} is MSO definable \cite{CouYuri,CouLMCS2020}.\ The case
of \textbf{IBO} was left as a conjecture.\ We will prove the following two results.

\bigskip

\textbf{Theorem 4.2 }: (1) The class \textbf{IBO}\ is effectively MSO$_{fin}$ definable.\ 

(2) This class is uFO definable.

\bigskip

Assertion (2) is not effective: we do not know the defining sentence.\ To the
opposite, Assertion (1) is.\ It entails that the class of bounds of
\textbf{IBO} is MSO definable (among finite structures).\ One can prove that
$Bnd($\textbf{IBO}) is finite, but this fact and the knowledge of the defining
MSO-sentence are not sufficient to yield an algorithm (see Section 6).

\bigskip

We will consider ternary structures $(N,B)$ that always satisfy the uFO
expressible properties A1-A6.\ These properties hold in every structure in
\textbf{IBO}\ but do not characterize this class (Proposition 3.22 of
\cite{CouLMCS2020}).

\subsection{Preliminary results on IBO}

\textbf{Defintion 4.3} : The \emph{Gaifman graph} of a ternary structure
$S=(N,B)$ is the graph $\mathit{Gf}(S)$ whose vertex set is $N$\ and that has
an edge $u-v$ if and only if $u$ and $v$ belong to some triple in $B$. We say
that $S$ is \emph{connected} if $\mathit{Gf}(S)$ is.\ If it is not, then $S$
is the disjoint union of the induced structures $S[X]$ for all connected
components $\mathit{Gf}(S)[X]$\ of $\mathit{Gf}(S)$.

\bigskip

\textbf{Lemma\ 4.4:} (1) A structure $S$ is in \textbf{IBO} if and only if its
connected components are.

(2) If a structure $S$ in \textbf{IBO} is connected, then it is an induced
betweenness structure of an O-tree.\ 

\textbf{Proof: }(1) The "only if" direction is clear by the definitions.
Conversely, assume that each connected component of a ternary structure
$S=(N,B)$ is in \textbf{IBO}. For each of them $S[X]$, let $U_{X}:=(M_{X}%
,\leq_{X})$\ be a defining O-forest (we have $M_{X}\supseteq X)$.\ We let
$\mathbb{N}^{R}$ be $\mathbb{N}$ ordered by reversing the natural order.\ We
assume these forests $U_{X}$ pairwise disjoint and disjoint from
$\mathbb{N}^{R}.$ \ We let $W$ be the union of $\mathbb{N}^{R}$ and the
$U_{X}$'s that we order as follows :

\begin{quote}
$x\leq_{W}y$ if and only if

$x\leq y$ in $\mathbb{N}^{R}$ or $x\leq_{X}y$ for some component $X$, or $x$
is in some $M_{X}$ and $y\in\mathbb{N}$.
\end{quote}

Then $W$ is an O-tree and $B=B_{W}\cap N^{3}.$

\bigskip

(2) Let $S=(N,B)$ be such that $B=B_{U}\cap N^{3}$ for some O-forest
$U=(M,\leq)$.\ Let $M^{\prime}$ be the union of the lines $L_{\geq}(x)$ of $U$
for all $x\in N$. Then $U^{\prime}:=U[M^{\prime}]$ is an O-forest and
$B=B_{U^{\prime}}\cap N^{3}$.\ We prove that it is an O-tree if furthermore
$S$ is connected.\ If $x$ and $y$ belong to a triple in $B$, then they have an
upper-bound in $M^{\prime}$\ by the definition of $B_{U}$\ and, furthermore,
any $x^{\prime}\geq x$ and $y^{\prime}\geq y$ also have an upper-bound in
$M^{\prime}$. Let $u,v\in M^{\prime}$.\ There is a path $x_{1}-x_{2}%
-...-x_{n}$ in $\mathit{Gf}(S)$ such that $u\geq x_{1}$ and $v\geq x_{n}.$
Hence we have $z_{1},z_{2},...,z_{n-1}$ such that:

\begin{quote}
$z_{1}$ is an upper-bound of $u$ and $x_{2}$,

$z_{2}$ is an upper-bound of $z_{1}$ and $x_{3}$, ..., and finally

$z_{n-1}$ is an upper-bound of $z_{n-2}$ and $v\geq x_{n}.$
\end{quote}

We have $z_{n-1}\geq u$.\ Hence, $U^{\prime}$ is an O-tree.$\square$

\bigskip

The converse of Assertion (2) may be false: consider a \emph{star}
$T=(N,\leq)$ with root\footnote{All other nodes are adjacent to the root.} $r$
and $S:=(N-\{r\},B)$ where $B:=B_{T}[(N-\{r\})].$ Then, $S$ is in
\textbf{IBO}, defined from a tree, but not connected as $B$ is empty.

\bigskip

\textbf{Definition 4.5\ }: \emph{Marked join-trees and related notions}

(a) A \emph{marked join-tree }is a 4-tuple $T=(M,\leq,M_{\oplus},M_{\otimes})$
such that $(M,\leq)$ is a join-tree and $M_{\oplus},M_{\otimes}$ are disjoint
subsets of $M$ that contain no leaf. We let $V_{T}:=M-(M_{\oplus}\uplus
M_{\otimes}).$ Its \emph{size} is defined as $\left\vert M\right\vert $, the
cardinality of $M$.

(b) We define the \emph{betweenness relation} $B_{T}\subseteq V_{T}^{3}$ of
$T$ as follows:

\begin{quote}
$B_{T}(x,y,z):\Longleftrightarrow$ \ \ $\neq(x,y,z)\wedge x\sqcup_{T}z\notin
M_{\oplus}~\wedge$

$\qquad\qquad\qquad\qquad([x<y\leq x\sqcup_{T}z]\vee\lbrack z<y\leq
x\sqcup_{T}z]).$
\end{quote}

The join $x\sqcup_{T}z$ is always defined as $T$ is a join-tree.\ We have
$B_{T}(x,y,z)$ if $x<y<z$.

We define the \emph{betweenness structure} of $T$ as $S_{T}:=(V_{T},B_{T}%
)$.\ Its Gaifman graph has vertex set $V_{T}$.

(c) If we delete from $T$\ all nodes of degree 1 belonging to $M_{\oplus
}\uplus M_{\otimes}$, we obtain a marked join-tree having the same betweenness
structure and that join-embeds into $T$ (cf. Definition 3.1(e)). We call
\emph{reduced} such a marked join-tree.

(d) If $M_{\oplus}$\ is empty, then $(V_{T},B_{T})\in$ \ \textbf{IBQT}.

(e) We say that a marked join-tree $U=(N,\leq,N_{\oplus},N_{\otimes})$
join-embeds into a marked join-tree $T=(M,\leq,M_{\oplus},M_{\otimes})$ if
there is a join-embedding of $(N,\leq)$ into $(M,\leq)$ that maps $N_{\oplus}$
to $M_{\oplus}$ and $N_{\otimes}$ to $M_{\otimes}.$

\bigskip

\textbf{Lemma 4.6\ }: Let $T=(M,\leq,M_{\oplus},M_{\otimes})$ be a marked join-tree.\ 

(1) If $U=(N,\leq,N_{\oplus},N_{\otimes})$ join-embeds into $T$, then
$B_{U}\ =B_{T}[N\cap V_{T}].$

(2) If $X\subseteq V_{T}$ and $B=B_{T}[X],$ then there exists $U$ as in (1)
such that $B_{U}=B$.

(3) If $T_{1},...,T_{n},..$.\ is a sequence of marked join-trees such that
$T_{n}\subseteq_{j}T_{n+1}$ and $T$ is the union of the $T_{n}$'s, then
$B_{T}$\ is the union of the increasing sequence $B_{T_{1}}\subseteq
_{i}B_{T_{2}}\subseteq_{i}...B_{T_{n}}\subseteq_{i}...$

\textbf{Proof }: (1) Since $U$ join-embeds into $T$, if $x,y\in N\cap V_{T}$,
then $x\sqcup_{U}y=x\sqcup_{T}y$ and this join belongs to $M_{\oplus}$
(resp.\ $M_{\otimes}$ ) if and only if it belongs to $N_{\oplus}$ (resp.
to\ $N_{\otimes}$). The result follows from the definitions.

(2) Let $T=(M,\leq,M_{\oplus},M_{\otimes})$ be a marked join-tree and
$N\subseteq M$. Let us remove from $T$ all subtrees $T/u$ that contain no node
of $N$.\ We obtain $U=(N^{\prime},\leq,N_{\oplus},N_{\otimes})$, a marked
join-tree that join-embeds into $T$ and $V_{U}=N\cap V_{T}.$ Hence we have
$B=B_{T}[N]=B_{U}\ $ by (1).

(3) We have $T_{n}\subseteq_{j}T$ for each $n$. The result follows.
$\ \ \ \square$

\bigskip

\textbf{Proposition 4.7} : (1) A structure $S=(N,B)$ is\ in \textbf{IBO} if
and only if $B=B_{T}$\ for a marked join-tree $T=(M,\leq,M_{\oplus}%
,M_{\otimes})$ such that $V_{T}=N$.\ 

(2) If $N$ is finite, then $T$ can be chosen finite of size at
most\ $2\left\vert N\right\vert -1$.

\textbf{Proof: }(1) \textbf{ }"If"\textbf{ }direction.\textbf{\ }Let\textbf{ }
$S_{T}:=(V_{T},B_{T})$ be defined from a marked join-tree $T=(M,\leq
,M_{\oplus},M_{\otimes}).\ $We will construct an O-tree $U=(W,\leq^{\prime})$
such that $M-M_{\oplus}\ \subseteq W$\ and $B_{U}[V_{T}]=B_{T}.$

For each node $x$ in $M_{\oplus}$, we let $\mathbb{N}_{x}^{R}$ be an
isomorphic copy of $\mathbb{N}$ ordered by reversing the natural
ordering.\ Hence $\mathbb{N}_{x}^{R}$ has no least element.\ We choose these
copies pairwise disjoint and disjoint with $M$.\ 

We define $U:=T[x\longleftarrow\mathbb{N}_{x}^{R};x\in M_{\oplus}]$.\ It is an
O-tree by Definition 3.2 (where substitutions are defined).

If $x\sqcup_{T}z\in M_{\oplus}$, then $x$ and $z$ have no join in $U$.

Let $x,y,z\in V_{T}$\ be such that $B_{T}(x,y,z)$ holds.\ If $x<y<z$ or
$z<y<x$ in $T,$ then the same holds in $U$\ and $B_{U}(x,y,z)$ holds.
Otherwise, $x$ and $z$ are incomparable and $x<y\leq x\sqcup_{T}z>z$ or
$x<x\sqcup_{T}z\geq y>z.\ $Then, $x\sqcup_{T}z$ is either in $V_{T}$ or is
labelled by $\otimes$. In both cases, $x\sqcup_{T}z$ is the join of $x$ and
$z$ in $U$.\ Hence, $B_{U}(x,y,z)$ holds.\ 

Conversely, assume that $x,y,z\in V_{T}$\ and $B_{U}(x,y,z)$ holds. If $x<y<z$
or $z<y<x$ in $U,$ then the same holds in $T$\ and $B_{T}(x,y,z)$
holds.\ Otherwise, $x$ and $z$ are incomparable\ and $x<y\leq x\sqcup_{U}z>z$
or $x<x\sqcup_{U}z\geq y>z$.\ Then $x$ and $z$ have a join $m$ in $T$.\ It
must be in $V_{T}\cup M_{\otimes}$, otherwise, $x\sqcup_{U}z$ does not exist
because it would be the minimal element of $\mathbb{N}_{m}^{R}.$ Hence
$B_{T}(x,y,z)$ holds. Hence $S\in~$\textbf{IBO}.

"Only if" direction.\ \ Conversely, assume that $S=(N,B)$ in\textbf{ IBO} is
defined from an O-tree $U=(M,\leq)$ such that $N\subseteq M$ and $B=B_{U}\cap
N^{3}.$ We can assume that for every $y\in M$, we have $x\leq y$ for some
$x\in N$: if this is not the case, we replace $M$ by the union $M^{\prime}$ of
the upwards closed lines $L_{U\geq}(x)$ for all $x\in N$ \ and, letting
$U^{\prime}:=(M^{\prime},\leq),$ we have $N\subseteq M^{\prime}$ and
$B=B_{U^{\prime}}\cap N^{3}.$

Let $W=(P,\leq)$ be the join-completion of $U$, cf Definition 3.3.\ We label
by $\otimes$ a node in $M-N$, and by $\oplus$ a node in $P-M$.\ These latter
nodes have been added to $U$ in place of missing joins, according to
Proposition 3.4.

\emph{Claim }: $B=B_{W}$.

\emph{Proof} : $B\subseteq B_{W}$. Let $B(x,y,z)$.\ If $x<y<z$ or $z<y<x$ in
$U$ then the same holds in $W$\ and $B_{W}(x,y,z)$ \ holds.

Otherwise, $x$ and $z$ are incomparable\ and $x<y\leq x\sqcup z>z$ or
$x<x\sqcup z\geq y>z$ in $U$.\ Then $x\sqcup z$\ is in $N$ or is labelled by
$\otimes$ in $W$. Hence, $B_{W}(x,y,z)$ holds. Then $x\sqcup z$ is not
labelled by $\oplus.$

Conversely, assume that $B_{W}(x,y,z)$ holds.\ A similar proof establishes
that $B(x,y,z)$ holds. $\ \ \ \square$

\bigskip

If $S=(N,B)$ in\textbf{ IBO} is defined from an O-forest $U=(M,\leq)$ as
opposed to an O-tree, then its connected components are defined by
O-trees.\ For each of them, we have a marked join-tree.\ We put them together
in a marked join-tree with a root labelled by $\oplus$. (Similarly to the
proof of Lemma 4.4(2)).

\bigskip

(2) Let $S=(N,B)$ in \textbf{IBO} be finite and defined from a marked
join-tree $T=(M,\leq,M_{\oplus},M_{\otimes})$ such that $N=V_{T}$ and
$B=B_{T}.$ By removing the nodes in $M_{\oplus}\cup M_{\otimes}$ of degree 1,
we obtain a reduced marked join-tree that defines $S$\ and has at
most\ $2\left\vert N\right\vert -1$ nodes.$\ \square$

\bigskip

\textbf{Remark 4.8 : }We observed in Proposition 2.15 of \cite{CouLMCS2020}
that a finite structure in \textbf{IBO}\ may not be defined from any
\emph{finite} O-forest $U$\ (cf.\ Definition 4.1). Marked join-trees remedy
this "defect" and yield Proposition 4.7, a key fact for our proof.

\bigskip

\textbf{Example 4.9}\ : We consider $S=(N,B)$ in \textbf{IBO}\ defined from
the infinite O-tree on the left of Figure 1 where $N=\{0,a,a^{\prime
},b,b^{\prime},$ $c,c^{\prime},d,d^{\prime},e,e^{\prime}\}$ and the dotted
line represents $\mathbb{N}^{R}$, without a least node.\ We have:

\begin{quote}
(a) $B(a^{\prime},a,0),B(b^{\prime},b,0),B(c^{\prime},c,0),B(d^{\prime
},d,0),B(e^{\prime},e,0),$

(b) $B^{+}(a^{\prime},a,b,b^{\prime})$, $B^{+}(c^{\prime},c,d,d^{\prime})$,
$B^{+}(a^{\prime},a,e,e^{\prime}),B^{+}(b^{\prime},b,e,e^{\prime}),$

$B^{+}(c^{\prime},c,e,e^{\prime})$ and $B^{+}(d^{\prime},d,e,e^{\prime})$.
\end{quote}

We do not have $B^{+}(a^{\prime},a,c,c^{\prime})$ because $a$ and $c$ have no join.

The right part shows a finite marked join-tree $T=(M,\leq,M_{\oplus
},M_{\otimes})$\ where $z$ has been added as join of $x$ and $y$ and the nodes
$2,3,...,n,...$\ of degree 1 above $z$\ have been deleted (cf.\ Definition 3.1
for the degree).

We have $M=N\cup\{1,x,y,z\},M_{\oplus}=\{z\},M_{\otimes}=\{1,x,y\}$ and
$B_{T}=B$. \ $\square$

\bigskip%

\begin{figure}
[ptb]
\begin{center}
\includegraphics[
height=2.5512in,
width=3.1081in
]%
{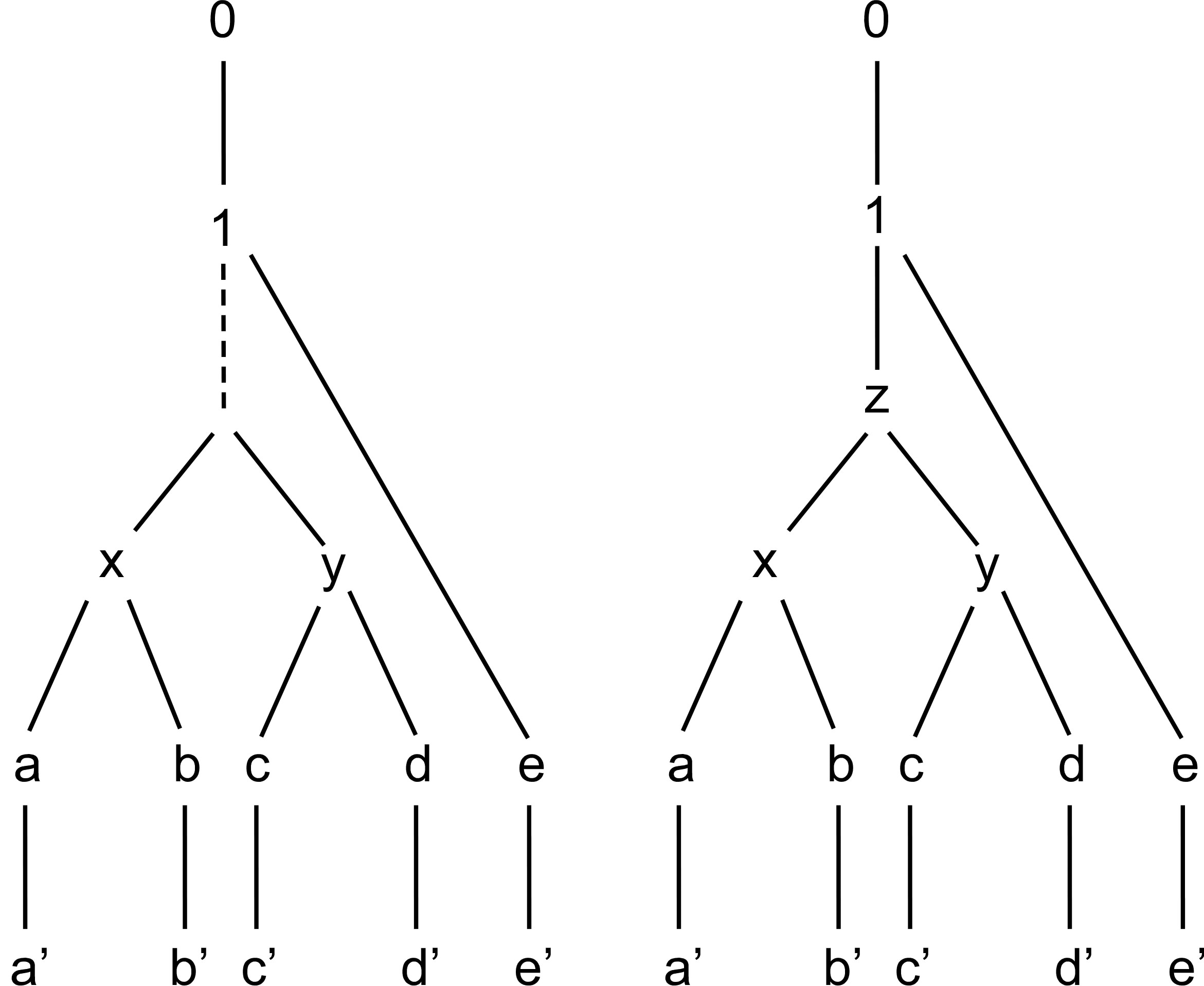}%
\caption{See Example 4.9.\ In the O-tree to the left, we have
$0>1>2>3>...>n>...$ above $x$ and $y$.}%
\end{center}
\end{figure}

\bigskip

\textbf{Proposition 4.10 }: The class \textbf{IBO} is finitary, that is, $S$
is in \textbf{IBO} if and only if each of its finite induced substructures is.

\textbf{Proof :} The "only if" direction is clear as, by its definition, the
class \textbf{IBO}\ is hereditary, \emph{i.e.}, closed under taking induced substructures.

"If" direction. First, some observations.\ If $S=(V_{T},B)$ is defined from a
marked join-tree $T=(N,\leq,N_{\oplus},N_{\otimes})$ and $S^{\prime}%
\subseteq_{i}S$, then the restriction $T^{\prime}$ of $T$ to $\{x\in N\mid
x\geq y$ \ for some $y\in V_{T}\}$ \ is a marked join-tree that defines
$S^{\prime}$.\ By reducing it (Definition 4.5(c)), we get $T^{\prime\prime
}\subseteq_{j}T$ that defines $S^{\prime}.$

By Proposition 4.7,\ each finite structure $S=(N,B)$ in \textbf{IBO}\ of size
$m=\left\vert N\right\vert $ is defined by marked join-trees of size at most
$2m-1$. We let $J(S)$ be the finite set of all such join-trees, up to isomorphism.

For proving the statement, we let $S=(N,B)$ be infinite. It is the union of an
increasing sequence $S_{1}\subset_{i}S_{2}\subset_{i}...\subset_{i}%
S_{n}\subset_{i}...$ of finite induced substructures that we assume to be in
\textbf{IBO}.

We will use the following version of Koenig's Lemma.\ Let $A_{1}%
,A_{2},...,A_{n}$,... be an infinite sequence of pairwise disjoint finite
sets, and $A$ be their union. Let $R\subseteq A\times A$ be such that for
every $b$ in $A_{n},n>1$, there is $a\in A_{n-1}$ such that $(a,b)\in
R$.\ Then, there exists an infinite sequence $a_{1},...,a_{n},...$\ such that
$(a_{n-1},a_{n})\in R$ for each $n>1$.

The finite sets $J(S_{n})$ are pairwise disjoint.\ We define

$R:=\{(T,T^{\prime})\mid T\in J(S_{n-1}),T^{\prime}\in J(S_{n}),n>1$ and
$T\subset_{j}T^{\prime}\}.$

It follows from Lemma 4.6(2)\ \ that if $T^{\prime}\in J(S_{n})$ and $n>1,$ we
have $(T,T^{\prime})\in R$\ for some $T\in J(S_{n-1}).$

Hence, there is an infinite sequence of marked join-trees trees

$T_{1}\subset_{j}T_{2}\subset_{j}...\subset_{j}T_{n}\subset_{j}...$ such that
$T_{n}\in J(S_{n})$ for each $n$.\ 

By Lemma 4.6(3), their union is a marked join-tree $T$ such that $T_{n}%
\subset_{j}T$\ for each $n$. We obtain an increasing sequence of finite marked
join-trees whose union is a marked join-tree that defines $S$.\ Hence $S\in
$\textbf{ IBO. }$\square$

\bigskip

The proof of Theorem 4.2(1) reduces to that of the following proposition.

\bigskip

\textbf{Proposition 4.11}: There is an MSO-sentence that characterizes the
finite connected structures in \textbf{IBO}\ among the finite ternary structures.

\bigskip

\textbf{Proof of Theorem 4.2(1), }assuming proved Proposition 4.11\textbf{\ }:
\textbf{ }

Let $\varphi$ be an MSO-sentence such that, for every finite ternary structure
$S=(N,B)$ :

\begin{quote}
$S\models\varphi$ if and only if $S$ is connected and belongs to \textbf{IBO}.
\end{quote}

Consider the MSO$_{fin}$ sentence $\psi:$

\begin{quote}
$\forall X.(\gamma(X)\wedge Fin(X)\Longrightarrow\varphi\lbrack X]),$
\end{quote}

where $\gamma(X)$ expresses that $X$ is connected in the Gaifman graph
$\mathit{Gf}(S)$ and $\varphi\lbrack X]$ is the relativization of $\varphi$ to
$X$.

\emph{Relativizing} a sentence to a set, here $X$, is a classical construction
in logic, see \emph{e.g.} \cite{CouEng}, Section 5.2.1 for monadic
second-order logic. If $S=(N,B)$ is a ternary structure and $X\subseteq N$,
then $S\models\varphi\lbrack X]$ if and only if \ $S[X]\models\varphi$.

We prove that $S\models$ $\psi$ if and only if $S$ is in \textbf{IBO}.

If $S$ is in \textbf{IBO}, then every induced substructure $S[X]$, in
particular every finite and connected one satisfies $\varphi$, hence
$S\models$ $\psi$.

Conversely, assume that $S\models$ $\psi$.\ \ Let $X$\ be a finite subset of
$N$.\ If it is connected in $\mathit{Gf}(S)$, then $\varphi\lbrack X]$ holds
hence \ $S[X]$ is in \textbf{IBO}.\ Otherwise, it is a disjoint union of
connected sets in $\mathit{Gf}(S)$.\ For each of them, say $Y$, the validity
of $\psi$ \ implies that $\varphi\lbrack Y]$ holds, $S[Y]$ is in \textbf{IBO}
and so are\ $S[X]$ by Lemma 4.4(1) and $S$ by Proposition 4.10. $\ \square$

\subsection{Proof of Proposition 4.11\ }

Proposition 4.11 is the main technical result.\ We will only handle finite
objects: graphs, rooted trees, rooted forests and structures $(N,B)$. All
trees and forests will be rooted, defined as partial orders $(N,\leq)$ and
simply called trees and forests. We need some more definitions.

\bigskip

\textbf{Definition 4.12\ }: \emph{Forests compatible with a ternary relation.}

A rooted forest $T=(N,\leq_{T})$ is\emph{ compatible with a relation}
$B\subseteq N^{3}$ satisfying Axioms A1-A6 (Definition 4.1) if, for all
$x,y,z\in N$ :

\begin{quote}
(i) if $B(x,y,z)$ holds, then $x<_{T}y$ or $z<_{T}y$,

(ii) if $B(x,y,z)$ and $x<_{T}y>_{T}z$ hold, then $y=x\sqcup_{T}z.$

(iii) if $x<_{T}z$, then $B(x,y,z)$ holds if and only if $x<_{T}y<_{T}z$.
$\square$
\end{quote}

\bigskip

\textbf{Lemma 4.13\ }: Let $S=(N,B)\in$\textbf{ IBO} be finite, connected and
defined from a finite reduced marked tree\footnote{Every finite tree is a
join-tree.} $U=(N\uplus N_{\oplus}\uplus N_{\otimes},$ $\leq_{U},N_{\oplus
},N_{\otimes})$.

(1) Then $T:=U[N]=(N,\leq_{T})$ is a finite forest compatible with $B$, where
$\leq_{T}$ is the restriction of $\leq_{U}$ to $N$.

(2) The order $\leq_{T}$ is FO definable in the structure $(N,B,R)$ where $R$
is the set of roots of $T$, \emph{i.e.} of maximal elements with respect to
$\leq_{T}$. $\square$

\bigskip

The forest $T$\ is not necessarily a tree because the root of $U$ need not be
in $N$. This root cannot be labelled by $\oplus$, otherwise $S$ is not
connected (we exclude the trivial case where $N$ is singleton).

\textbf{Proof} : (1) Let $S,T,U$ as in the statement.\ \ 

(i) If $B(x,y,z)$ holds, then:

\begin{quote}
either $x<_{U}y<_{U}z$ or $z$ $<_{U}y<_{U}x,$

or $x\bot_{U}z\wedge\lbrack(x<_{U}y\leq_{U}x\sqcup_{U}z)\ \vee\ (z<_{U}%
y\leq_{U}x\sqcup_{U}z)],$

where in the latter case, $x\sqcup_{U}z\in N\cup N_{\otimes}.$
\end{quote}

In all cases, we have $x<_{U}y$ or $z<_{U}y$,\ hence $x<_{T}y$ or $z<_{T}y$.

(ii) If $B(x,y,z)$ and $x<_{T}y>_{T}z$ hold, then the above description of
$B(x,y,z)$ shows that $y\leq_{U}x\sqcup_{U}z.$ As we have $x<_{U}y>_{U}z,$ we
must have $y=x\sqcup_{U}z$.\ If $y$ is not $x\sqcup_{T}z$, we have $m\in N$
such that $x<m$ and $z<m<y$ in $T$ and in $U$.\ \ But then $y$ is not the join
of $x$ and $z$ in $U$ and $y=x\sqcup_{T}z$.

(iii) Clear from the definitions because $\leq_{T}$ is the restriction of
$\leq_{U}$ to $N$. \ \ 

\bigskip

(2) \ If $R=\{r\}$, then $x\leq_{T}y$ if and only if $x=y$ or $y=r$ or
$B(x,y,r)$ holds.

Otherwise, the root of $U$ is in $N_{\otimes}$ and has degree at least 2. Let
$x$ and $y$ be not in $R$.

\emph{Claim}\ : (a) If $r\in R$, we have $x<_{T}r$ if and only if
$B(x,r,r^{\prime})$ holds for some $r^{\prime}\in R$.

(b) We have $x<_{T}y$ if and only if $B(x,y,r)$ holds for some $r\in R$.

\emph{Proof}: (a) Assume that $x<_{T}r$.\ There is $r^{\prime}\in R$\ such
that $r\sqcup_{U}r^{\prime}$ has label $\otimes$.\ Hence $B(x,r,r^{\prime})$ holds.

Conversely, if $B(x,r,r^{\prime})$ holds for some $r^{\prime}\in R,$ we have
$x<_{T}r$ or $r^{\prime}<_{T}r$ because $T$ is compatible with $B$.\ As $r$
and $r^{\prime}$ are different and are distinct roots of $T$, they are
incomparable and we have $x<_{T}r$.\ 

(b) If $x<_{T}y$, we have $x<_{T}y<_{T}r$ for some $r\in R$.\ Hence $B(x,y,r)$
holds since $T$ is compatible with $B.$

Conversely, if $B(x,y,r)$ holds for some $r\in R$, then, we have $x<_{T}y$ or
$r<_{T}y$.\ The latter is not possible as $r$ is a root.\ \ \ $\square$

\bigskip

Let $\psi(R,x,y)$ be the following FO formula (an FO\ formula may have free
set variables and use atomic formulas $x\in X$):

\begin{quote}
$x=y\vee\lbrack x\neq y\wedge\exists r.(R=\{r\}\wedge\lbrack y=r\vee
B(x,y,r)])]$

$\qquad\vee\lbrack x\neq y\wedge\exists r,r^{\prime}\in R.(y=r\wedge
B(x,y,r^{\prime}))]$

$\qquad\vee\lbrack x\neq y\wedge y\notin R\wedge\exists r\in R.B(x,y,r)].$
\end{quote}

By the claim, it defines $x\leq_{T}y$ since $R$\ is the set of roots of $T$. \ \ \ 

We let $\varphi(R)$ be the FO\ formula relative to ternary structures
$S=(N,B)$ expressing the following:

\begin{quote}
"$R\subseteq N$, the binary relation $x\leq y$ on $N$ defined by $S\models
\psi(R,x,y)$ is a partial order and $T:=(N,\leq)$ is a forest that is
compatible with $B$ and whose set of roots is $R$". $\square$
\end{quote}

\bigskip

\textbf{Proposition 4.14 :} Let $S=(N,B)$ be finite and satisfy properties A1-A6.\ 

(1) For every $R\subseteq N$ such that $S\models\varphi(R)$, if we let $\leq$
be defined by $\psi(R,x,y)$, then $T:=(N,\leq)$ is a forest compatible with
$B$.

(2) Every forest $T:=U[N]$ defined from a finite marked tree $U=(N\uplus
N_{\oplus}\uplus N_{\otimes},$ $\leq_{U},N_{\oplus},N_{\otimes})$\ such that
$B_{U}=B$ is described by the formulas $\varphi(R)$ and $\psi(R,x,y)$.

\textbf{Proof }: The first assertion follows from the definition of $\varphi$.
The second one follows from Lemma 4.13(2). $\ \ \square$

\bigskip

All forests $T$\ compatible with $B$\ of potential interest for checking that
$S$ is in \textbf{IBO}\ can be described in terms of their sets of roots $R$
by the existential MSO-formulas \ $\exists R.\varphi(R)$\ $\ $and
$\psi(R,x,y)$.

We will construct MSO-formulas to "check" that a "guessed" forest $T$
satisfies additional requirements implying that $T:=U[N]$ for some finite
marked tree $U$ witnessing that $S\in$\textbf{IBO}.

In some $T$ that has been "guessed", we will insert (if possible) finitely
many nodes labelled by $\oplus$ or $\otimes,$ so as to make it into the
desired marked tree $U$. We will insert nodes in $T$\ in the following cases:

(1) If $a^{\prime}<a$, $b\bot a$, $B(a^{\prime},a,b)$ holds and there is no
$x$ in $T$\ such that $\{a,b\}<x$ and $B(a,x,b)$ holds, then we insert
$a\sqcup_{U}b$ labelled by $\otimes$ such that $a\sqcup_{U}b<m$ where $m$ is
any upper-bound in $T$\ of $\{a,b\}$.

(2) If $a^{\prime}<a,b\bot a$ but $B(a^{\prime},a,b)$ does not hold, then we
insert $a\sqcup_{U}b$ as above labelled by $\oplus$.

In Case (1) $a$ and $b$ may have a join $m$ in $T$ but we need to insert a
"new join" $a\sqcup_{U}b<m$.\ 

\bigskip

\textbf{Example 4.15\ }: The left part of Figure 2\ shows a tree
$T$\ constructed from the following facts relative to $N:=\{a^{\prime
},a,b,c^{\prime},c,d,e^{\prime},e,1,0\}:$

\begin{quote}
(a) $B^{+}(a^{\prime},a,1,0),B(b,1,0),B^{+}(c^{\prime},c,1,0),B(d,1,0),B^{+}%
(e^{\prime},e,1,0),$

(b) $B^{+}(a^{\prime},a,c,c^{\prime})$, $B(a^{\prime},a,d),B(b,c,c^{\prime}),$

(c) $B^{+}(a^{\prime},a,1,e,e^{\prime}),B^{+}(b,1,e,e^{\prime}),B^{+}%
(c^{\prime},c,1,e,e^{\prime})$ and $B^{+}(d,1,e,e^{\prime})$.
\end{quote}

Facts (b) indicate the need of joins $a\sqcup c$, $a\sqcup d,$ and $b\sqcup c$
labelled by $\otimes$ in the marked tree $U$ to be constructed. These joins
are all equal to $x$\ in the tree in the middle of Figure 2.\ The absence of
facts $B(a^{\prime},a,b)$ and $B(c^{\prime},c,d)$ indicates the need of
$\oplus$-labelled joins $y$ and $z,$ respectively between $a$,$b$ and $x$, and
between $c$,$d$ and $x$.\ However, no triple in $B$\ necessitates that $b$ and
$d$ have a $\otimes$-labelled join.\ A corresponding marked tree is shown in
the middle of Figure 2.

Consider now $N^{\prime}:=\{a^{\prime},a,b,d,e^{\prime},e,1,0\}\subseteq N$
and $B^{\prime}:=B[N^{\prime}].$ We get a marked tree for $B^{\prime}$ by
deleting $c,c^{\prime}$ and $z$ from the previous one. However, another one is
shown to the right, that defines $(N^{\prime},B^{\prime})$ where $x$ is
labelled by $\otimes$ and $z$ by $\oplus$. \ \ $\square$%

\begin{figure}
[ptb]
\begin{center}
\includegraphics[
height=2.3973in,
width=4.3855in
]%
{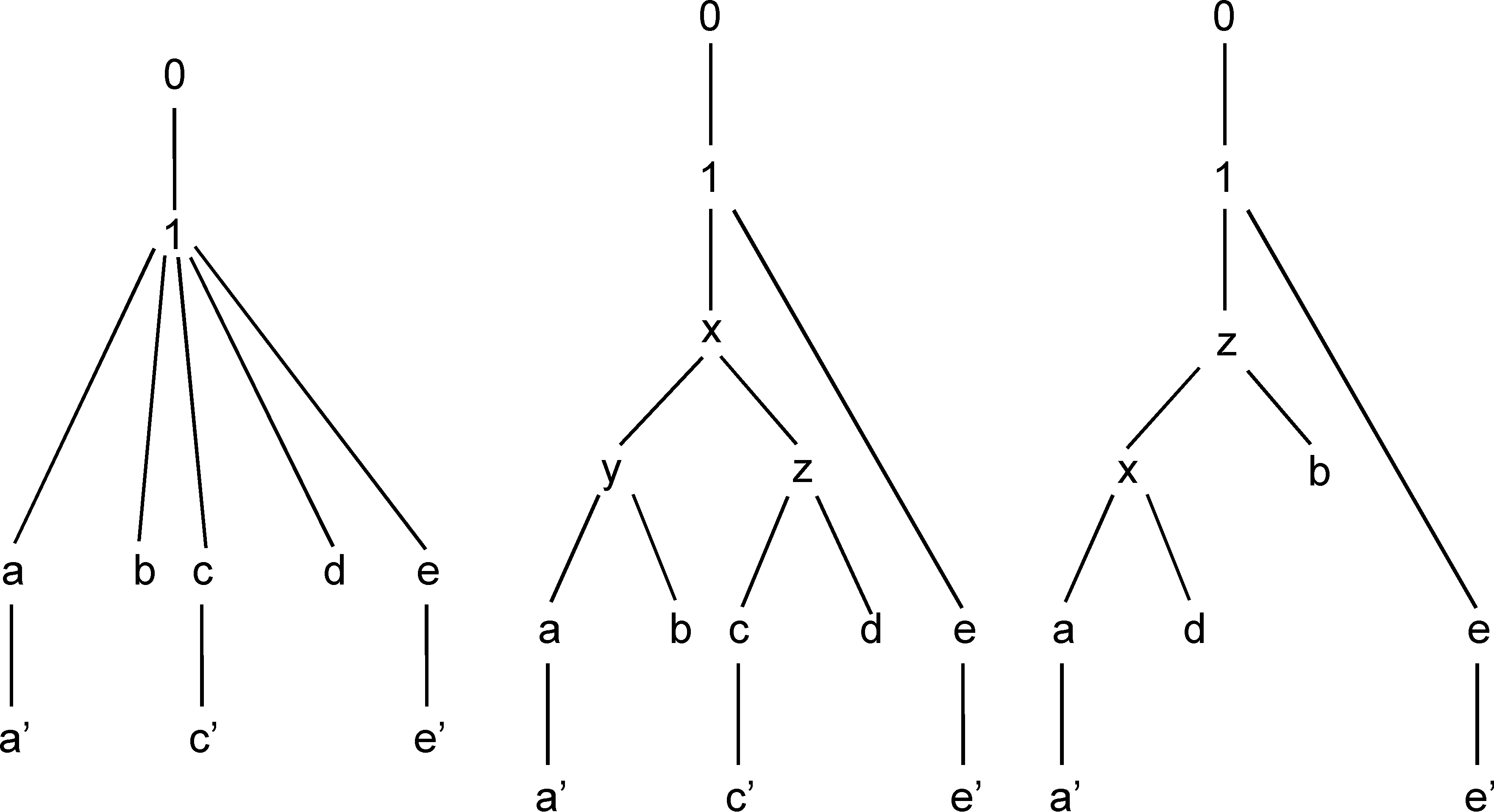}%
\caption{Example 4.15}%
\end{center}
\end{figure}

$\ \ $

We need more definitions. We let $B,T,U$ and$\ R$, be as in Lemma 4.13.

\bigskip

\textbf{Definitions 4.16\ : }\emph{Cographs and pp-cographs defined from
}$\emph{T}$\emph{ and, either }$\emph{U}$\emph{ or }$\emph{B}$\emph{.}

(a) For each node $x$ of $T$ with sons $y_{1},...,y_{s},s\geq2$, we define
$y_{i}\sim_{x}y_{j}$ if and only if $i=j\ $or $\ y_{i}\sqcup_{U}y_{j}\neq x$
so that this join has label $\otimes$ or $\oplus.$ It is clear that $\sim_{x}$
is an equivalence relation.

We have $y_{i}\sim_{x}y_{j}$ if and only if $B(y_{i},x,y_{j})$ does not hold,
by (ii) of compatibility, Definition 4.12.\ 

(b) For each class $C$\ of the equivalence relation $\sim_{x}$, we define
$G_{x,C}$ as the 2-graph $(C,E,C_{1},C_{2})$ such that:

\begin{quote}
$y\in C_{2}$ if and only if $y^{\prime}<y$ for some $y^{\prime}\in N$,

$y-z$ is an edge\footnote{$E$ denotes the set of edges.} if and only if $y$ or
$z$ is in $C_{2}$ and$\ y\sqcup_{U}z$ has label $\otimes.$
\end{quote}

There are no edges between vertices in $C_{1}$.\ \ Hence $G_{x,C}$ is a
pp-cograph.\ We obtain a cograph if we add edges $y-z$ such that $y$ and $z$
are in $C_{1}$ and$\ y\sqcup_{U}z$ has label $\otimes.$

(c) Let $R=\{r_{1},...,r_{p}\}$, $p\geq2$. We let $G_{root}$ be the 2-graph
$(R,E,R_{1},R_{2})$ defined as $G_{x,C}$ above, where $R$ replaces $C$ and
$y\in R_{2}$ if and only if $y^{\prime}<y$ for some $y^{\prime}\in N$. It is
also a pp-cograph. $\square$

\bigskip

\textbf{Lemma 4.17\ : } The edges $y-z$ of $G_{x,C}$ and $G_{root}$ are
characterized by the FO formula $(\exists y^{\prime}<y.B(y^{\prime}%
,y,z))\vee(\exists z^{\prime}<z.B(z^{\prime},z,y))$.

\textbf{Proof} : Consider $x\in N$ having sons $y$ and $z$ in a class $C$ of
$\sim_{x}.$

Let $y-z$ be an edge of $G_{x,C}$ such that $y\in C_{2}$ and$\ y\sqcup_{U}z$
has label $\otimes.$ Then $B_{U}(y^{\prime},y,z)$ holds for all $y^{\prime}<y$
and so does $B(y^{\prime},y,z)$ as $B=B_{U}$.

Conversely, if $y^{\prime}<y\wedge$ $B(y^{\prime},y,z)$ holds, then the join
$y\sqcup_{U}z$ must have label $\otimes$ or be in $N$. But in the latter case,
it must be $x$ as $y$ and $z$ are sons of $x.$ Hence, we have $B(y,x,z)$ but
then, we do not have $y\sim_{x}z.$\ Hence, $y-z$ is an edge of $G_{x,C}$.\ 

The proof is similar for $G_{root}$.\ The join $y\sqcup_{U}z$ cannot be in $N$
as $y,z$ are distinct roots. \ \ $\square$

\bigskip

It follows that the 2-graphs $G_{root}$ and $G_{x,C}$ can be defined from $B$
and $T$ only, without using $U$\ that we are actually looking for.
Furthermore, they can be described by FO\ formulas in the structure $(N,B,R)$.

\bigskip

The formulas $\varphi(R)$ and $\psi(R,x,y)$ are defined before Proposition 4.14.

\bigskip

\textbf{Proposition 4.18 }: Let $S=(N,B)$\ and $T=(N,\leq)\ $be defined by the
formula $\psi(R,x,y)$ from some $R$\ satisfying $\varphi(R)$.

(1) There exists a marked tree $U\supseteq T$ such that $B=B_{U}$ if and only
if the 2-graphs $G_{root}$ and $G_{x,C}$ are pp-cographs.

(2) This condition is FO\ expressible in the structure $(N,B,\leq).$

\textbf{Proof:} (1)\textbf{ }The \ "only if" direction follows from the
previous constructions.

Conversely, assume that each 2-graph $G_{x,C}$ (determined solely from $T$ and
$B$ by Lemma 4.17) as in the statement is a pp-cograph. By adding some edges
between its 1-vertices, we can get a cograph $H_{x,C}$ $\supseteq$ $G_{x,C}%
$.\ It is defined by an $\{\oplus,\otimes\}$-tree $t_{x,C}$ \ (Definition
2.1(c)), a tree whose internal nodes are labelled by $\oplus$ or $\otimes$ and
whose set of leaves is $C$.\ 

Similarly, if $T$ has several roots and $G_{root}$ is a pp-cograph, there is a
cograph $H_{root}$ $\supseteq$ $G_{root}$ defined by a $\{\oplus,\otimes
\}$-tree\ $t_{root}$ whose set of leaves is $R$.

By inserting in $T$ \ the internal nodes of $t_{x,C}$ between $x$ and the
nodes in $C$, for all relevant pairs $(C,x)$, and those of \ $t_{root}$ above
the roots of $T$, we get a marked tree $U$ such that $U\supseteq T=U[N]$ and
$B=B_{U}.$

This can be formalized as follows. By bottom-up induction, we define marked
trees $T_{x}$ and $T_{x,C}$ for each $x$ in $N$\ and equivalence class $C$\ of
the relation $\sim_{x}$. \ We assume that the trees $t_{x,C}$ and $t_{root}$
\ are pairwise disjoint.

(a) If $x$ is a leaf, then $T_{x}:=x$. There is no set $C$ to consider.

Otherwise, $T_{x}:=x(...,T_{x,C},...)$ where the list covers all equivalence
classes $C$\ of $\sim_{x}$. (We use the linear notation of finite rooted trees
defined in Section 1).

(b) If $C=\{y\}$, then $T_{x,C}:=T_{y}.$

Otherwise, we use the $\{\oplus,\otimes\}$-tree $t_{x,C}$ to define
$T_{x,C}:=t_{x,C}[...,y\longleftarrow T_{y}/y,...]$,\ denoting the
simultaneous substitution in $t_{x,C}$ of $T_{y}$ for each $y\in C$ (it is a
leaf of $t_{x,C}$).

(c) To complete the construction, we define $U:=T_{r}$ if $T$ is a tree with
root $r\in N$. Otherwise, $U:=t_{root}[...,r\longleftarrow T_{r},...]$
\ denoting the substitution in $t_{root}$ of $T_{r}$ for each leaf $r\in R$
(it is a leaf of $t_{root}$).

It is clear that $U$ is a marked tree $(N\uplus N_{\oplus}\uplus N_{\otimes},$
$\leq_{U},N_{\oplus},N_{\otimes})$ and that $T=U[N]$.

\emph{Claim }: $B_{U}\ =B$.

\emph{Proof : }Note that $\leq_{T}$ is the restriction of $\leq_{U}$ to $N$.

If $x$ and $z$ are comparable, then $(x,y,z)\in B_{U}$ if and only if
$x<_{U}y<_{U}z$ if and only if $x<_{T}y<_{T}z$ if and only if $\ \ (x,y,z)\in
B$ \ since $T$ is compatible with $B$.

We now assume $x\bot z$ and $(x,y,z)\in B.$

Let $u:=x\sqcup_{U}z$. By the compatibility of $T$ with $B$ (point (ii)), we
have $x<_{T}y$ or $z<_{T}y$.

(a) If $u\in N$, then $u=x\sqcup_{T}z$. Again by compatibility (point (i)), we
do not have $u<_{T}y$.\ Hence, we have $x<_{T}y\leq_{T}u>_{T}z$ or
$z<_{T}y\leq_{T}u>_{T}x.$ The same inequalities hold with $\leq_{U\text{ }}%
$hence $(x,y,z)\in B_{U}.$

(b) Otherwise, $u$ has label $\oplus$ or $\otimes.$ Let $x^{\prime}$ be
maximal in $N$\ such that $x\leq x^{\prime}$ and $z^{\prime}$ be similar for
$z$.

(b.1) If $x^{\prime}$ and $z^{\prime}$ have no upper-bound in $T$, they are
distinct roots and $u$ is an internal node of $G_{root}$.

As noted above, we have $x<_{T}y$ or $z<_{T}y$. Assume the first,
\emph{w.l.o.g.}. Then $x<_{T}y\leq_{T}x^{\prime}$.

If $u$ has label $\otimes$ then $B_{U}(x,y,z)$ holds by the definition of
$B_{U}$.\ 

(If $u$ has label $\oplus$ then $B_{U}(x,y,z)$ does not hold, but the
definition of the edges of $G_{root}$ gives that $B_{U}(x,y,z)$ does not hold.)\ 

(b.2) If $x^{\prime}$ and $z^{\prime}$ have a least upper-bound $m$ in $T$,
then, $u<_{U}m$. We have two cases:

\emph{Case 1 }: $B(x,m,z)$ holds.\ We cannot have $y>_{T}m$, hence, we have
$x<_{T}y\leq_{T}m>_{T}z$ or $z<_{T}y\leq_{T}m>_{T}x.$ The same inequalities
hold with $\leq_{U\text{ }},$\ hence $(x,y,z)\in B_{U}.$

\emph{Case 2} : if $B(x,m,z)$ does not hold.\ Then $x^{\prime}\sim
_{m}z^{\prime}$ (we cannot have $B(x^{\prime},m,z^{\prime})$) and so
$x^{\prime}$ and $z^{\prime}$ belong to a same class $C$\ of $\sim_{m}.$ Then
we use the same argument as above with $G_{m,C}$ instead of $G_{root}$.

The proof that $B_{U}\ \subseteq B\ $\ is similar. \ \ \ $\square$

This completes the proof of Assertion (1).

\bigskip

(2) The following facts can be expressed in the structure $(N,\leq,B)$ such
that $S=(N,B)$\ satisfies A1-A6 and $T=(N,\leq)$ is a forest compatible with
$B$ by MSO-formulas that are easy to write explicitely:

$\alpha(R,R_{1},R_{2})$ : $R$ is the set of root of $T$, it is not singleton
and ($R_{1},R_{2})$ is its partition defined in Definition 4.16(c).

$\beta(x,y,z)$ : $y<x\wedge z<x\wedge y\sim_{x}z,$ ($y$ and $z$ are sons of
$x$ in $T$) cf.\ Definition 4.16(a).

$\gamma(x,C,C_{1},C_{2}):$ $C$ is a set of sons of $x$ and an equivalence
class of $\sim_{x},$ ($C_{1},C_{2})$ is its partition defined in Definition 4.16(b).

$\eta(R,y,z):y-z$ is an edge of $G_{root}$.

$\eta^{\prime}(x,C,y,z):y-z$ is an edge of $G_{x,C}$.

$\pi(R):$ $R$ is not singleton and $G_{root}$ is a pp-cograph (we use $\alpha$
and $\eta$).

$\pi^{\prime}(x,C,y,z):$ $G_{x,C}$ is well-defined and is a pp-cograph (we use
$\gamma$ and $\eta^{\prime}$).

It is MSO expressible in $(N,B,\leq)$ by Proposition 2.4(1) whether the
2-graphs $G_{root}$ and $G_{x,C}$ are all pp-cographs. The condition of
Assertion (1)\ is thus MSO expressible in the structure $(N,B,\leq)$ by an
MSO-sentence $\mu$. $\square$

\bigskip

\textbf{Proof of Proposition 4.11: }We prove that an MSO-sentence can
characterize the finite connected structures in \textbf{IBO}\ among the finite
ternary structures. There is an MSO-sentence $\chi$ expressing that a ternary
structure $S=(N,B)$ is connected and satisfies A1-A6. The sentence over
$S=(N,B)$ defined as $\exists R.(\varphi(R)\wedge\mu^{\prime}(R))$ where
$\mu^{\prime}$ translates $\mu$ (of Proposition 4.18(2)) by using
$\psi(R,x,y)$ to define $\leq$ expresses well that $S$ is in \textbf{IBO} by
Proposition 4.18(1). Hence, $\chi$ $\wedge$ $\exists R.(\varphi(R)\wedge
\mu^{\prime}(R))$ is the desired sentence. $\ \ \square$

\bigskip

\subsection{Well-quasi-orderings and\ finite sets of bounds}

\bigskip

We recall definitions and a result from \cite{Pou}.

\bigskip

\textbf{Definitions 4.19 :}

(a) If $\mathcal{C}$ is a hereditary class of finite structures $S=(N,R_{1}%
,...,R_{p})$, if $m$ is the maximal arity of a relation $R_{i}$, we denote by
$\mathcal{U}(\mathcal{C})$ the class of structures $(N,R_{1},...,R_{p},U_{1},$
$...,U_{2m-1})$\ \ for all $S$ in $\mathcal{C}$, where $U_{1},...,U_{2m-1}$
are unary relations, hence that denote subsets of $N$.

(b) The class $\mathcal{U}(\mathcal{C})$ is \emph{well-quasi-ordered
}(implicitely by induced inclusion) if, for every infinite sequence
$S_{1},S_{2},...$ of structures in this class, there are $n<m$ such that
$S_{n}$ is isomorphic to an induced substructure of $S_{m},$ which we denote
by $S_{n}\subseteq_{i\sim}S_{m}$. \ $\square$

\bigskip

With these hypotheses and notation, Corollary 2.4\ of \cite{Pou} (also
Theorem\footnote{Fra\"{\i}ss\'{e} states the result with $2m$ instead of
$2m-1$ but translates from French the proof by Pouzet.} 13.2.3 of \cite{Fra})
states the following. (The class of bounds $Bnd(\mathcal{C}$)\ is defined in
Section 1. Its finiteness is up to isomorphism.)

\bigskip

\textbf{Theorem 4.20}\ : If $\mathcal{C}$ is hereditary and $\mathcal{U}%
$($\mathcal{C}$) is well-quasi-ordered, then $Bnd(\mathcal{C}$) is finite.\ 

\bigskip

The structures in $\mathcal{U}$(\textbf{IBO}) are of the form $(N,B,U_{1}%
,...,U_{5})$ for $(N,B)$ in \textbf{IBO}. Let \textbf{T}\ be the set of finite
structures $T=(N\uplus N_{\oplus}\uplus N_{\otimes},\leq,N_{\oplus}%
,N_{\otimes},U_{1},...,U_{5})$ such that $(N\uplus N_{\oplus}\uplus
N_{\otimes},\leq,N_{\oplus},N_{\otimes})$ is a marked tree (Definition
4.5(a)), $U_{1},...,U_{5}$ are subsets of $N$ and $\left\vert N_{\oplus}\uplus
N_{\otimes}\right\vert <\left\vert N\right\vert .$

Proposition 4.7(2) shows that every $S$ in $\mathcal{U}$(\textbf{IBO}) is
defined from a marked tree $T$\ belonging to \textbf{T}.\ We denote then
$S=S(T)$.\ Precisely, $S(T)=(N,B_{T^{\prime}},U_{1},$ \ $...,U_{5})$ where
$T^{\prime}$ is the marked tree $T=(N\uplus N_{\oplus}\uplus N_{\otimes}%
,\leq,N_{\oplus},N_{\otimes}),$ cf.\ Definition 4.5(b).

\emph{Fact} : If $T$,$T^{\prime}$ are in \textbf{T} and $T\subseteq_{j\sim
}T^{\prime}$ , then $S(T)\subseteq_{i\sim}S(T^{\prime})$.

It is a corollary of Lemma 4.6(1).

\bigskip

\textbf{Proposition 4.21} : The class of finite structures in $\mathcal{U}%
$(\textbf{IBO}) is well-quasi-ordered.

\textbf{Proof : }Let $S_{1},S_{2},...$ be an infinite sequence of finite
structures in $\mathcal{U}$(\textbf{IBO}).\ For each $S_{n}$, we let $T_{n}$
in \textbf{T} be such that $S(T_{n})=S_{n}$.\ By Kruskal's Theorem,
$T_{n}\subseteq_{j\sim}T_{m}$ for some $n<m$.\ The above fact yields
\ $S_{n}\subseteq_{i\sim}S_{m}$. \ $\square$

\bigskip

\textbf{Theorem 4.2(2)} : The class \textbf{IBO}\ has finitely many bounds.\ 

\textbf{Proof} :\ The hereditary class of finite structures belonging to
\textbf{IBO}\ has finitely many bounds by Proposition 4.21\ and Theorem 4.20.
The result holds by Proposition 4.10. \ $\square$

\bigskip

\textbf{Remark 4.22} : We recall from Definition 4.1 that \textbf{IBQT} is the
class of induced betweenness of join-trees.\ It is a proper subclass of
\textbf{IBO}. The structures in \textbf{IBQT} are defined from
marked\ join-trees $T=(N\uplus N_{\oplus}\uplus N_{\otimes},\leq,N_{\oplus
},N_{\otimes})$ such that $N_{\oplus}$\ is empty (Definition 4.5(d)).\ The
proof of Theorem 4.2(2)\ shows that $Bnd$(\textbf{IBQT}) is finite, hence that
\ \textbf{IBQT} is FO definable, without constructing the defining
sentence.\ The FO definability of \textbf{IBQT} is Theorem 3.1\ of
\cite{CouLMCS2020}, where the defining FO-sentence is the conjunction of
Conditions A1-A6 and A8\ of Definition 4.1.\ 

\bigskip

In Section 6, we will explain why computing the bounds of \textbf{IBO} is even
harder than computing those of probe cographs.

\section{Clique-width and the bounds of probe cographs}

\bigskip

We first review clique-width, then, we discuss some properties of probe
cographs, in view of determining their bounds.

\bigskip

\textbf{Definition 5.1\ \ :}\emph{ Clique-width.}

(a) Graphs are built with the help of vertex labels (in addition to the labels
of 2-graphs).\ Each vertex has a \emph{label} in a set $L$. The nullary symbol
$\boldsymbol{a}(x)$ where $a\in L$, denotes the isolated vertex $x$ labelled
by $a$.\ The operations are the union\ $\oplus$ of disjoint graphs (it does
not modify labels), the unary operation $add_{a,b}$ for $a,b\in L$, $b\neq a$,
that adds to a graph an edge between each $a$-labelled vertex and each
$b$-labelled vertex (unless they are already adjacent), the unary operation
$relab_{a\rightarrow b}$ that changes every vertex label $a$ into $b.$ \ 

\bigskip

(b) A term over the above defined operations is \emph{well-formed} if no two
occurrences of nullary symbols denote the same vertex (so that the graphs
defined by two arguments of any operation\ $\oplus$\ are disjoint). We
call\ them the \emph{clique-width terms}. Each term $t$ denotes a vertex
labelled graph $\boldsymbol{val}(t)$ whose vertices are those specified by the
nullary symbols of $t$. Its \emph{width} is the number of labels that occur in
$t$. The \emph{clique-width} of a graph $G$ without labels from $L$ (but
possibly with labels from another set like $\{1,2\})$, denoted by $cwd(G),$ is
the least width of a term $t$\ that denotes some vertex labelling of $G$.

\bigskip

(c) Clique-width terms may contain redundancies: for example, we have
$add_{a,b}(add_{c,d}(add_{a,b}(G)))=add_{c,d}(add_{a,b}(G))$\ and
$relab_{a\rightarrow b}(relab_{a\rightarrow c}(G))=relab_{a\rightarrow c}(G)$
for every graph $G$.\ It follows that each graph of clique-width at most $k$
is defined by infinitely many terms written with a fixed set $L$ of $k$
labels. However, one can "normalize" these terms so as to avoid redundancies.
This is done in\ Proposition 2.121\ of \cite{CouEng}.\ Let us call
\emph{normal} such a term. Then, each graph of clique-width at most $k$ is
defined by finitely many normal terms using the labels in $L:=[k]$.
Furthermore, the set $N_{k}$ of normal terms with labels in $[k]$ is
recognizable by a finite automaton, see \cite{CouEng}. $\square$

\bigskip

\textbf{Proposition 5.2\ \ : }The maximal clique-width of a probe cograph is 4.

\textbf{Proof : }The upper-bound, observed in \cite{Dal+},\ is easy to
establish. The bound 4 is reached by the probe cograph defined by the term
(cf.\ Definition 2.3(c)):

\begin{quote}
$[\bullet_{1}(1)\oplus(\bullet_{1}(2)\otimes\bullet_{2}(7))\oplus(\bullet
_{1}(3)\otimes\bullet_{2}(8))]\otimes$

$\qquad\qquad\lbrack\bullet_{1}(4)\oplus(\bullet_{1}(5)\otimes\bullet
_{2}(9))\oplus(\bullet_{1}(6)\otimes\bullet_{2}(10))]$
\end{quote}

where the vertices 1,...,6 are 1-vertices\ and the vertices 7 to 10\ are
2-vertices.\ It has clique-width\footnote{The verification has been done by
using the software TRAG \cite{TRAG} that is accessible on-line. It is based on
\cite{HS}.} 4.\ $\square$\ 

\bigskip

\textbf{Proposition 5.3:} Apart from $P_{6}$, the finitely many bounds of
probe cographs have diameter at most 4\ and clique-width at most 8. They are
connected and MSO definable.

\textbf{Proof : }Since the class of probe cographs is closed under disjoint
union, its bounds are connected.

We have observed that $P_{6}$\ of diameter 5\ is a bound.\ Any other graph of
diameter at least 5 has an induced path $P_{6}$, hence is not a bound.\ 

If a graph has $G-x$ has clique-width $k$, then $G$ has clique-width at most
$2k$ \cite{Gur}. Hence, as probe cographs have clique-width at most 4, their
bounds have clique-width at most 8.

If $\mathcal{C}$ is a hereditary class of finite graphs, then its bounds form
the class:

\begin{quote}
$Bnd(\mathcal{C}):=\{G\mid G\notin\mathcal{C}$ and $G-x\in\mathcal{C}$ \ for
each vertex $x$ of $G$\}.
\end{quote}

If $\mathcal{C}$ is defined by an MSO-sentence $\theta,$ then $Bnd(\mathcal{C}%
)$ is defined by the MSO-sentence :

\begin{quote}
$\lnot\theta\wedge\forall X.($"$X$ is the set of all vertices minus one"
$\Longrightarrow\theta\lbrack X]$)\ \ \ 
\end{quote}

By Proposition 2.4(1) the class of pp-cographs is FO-definable.\ Hence, the
class of probe cographs is MSO-definable: an existential set quantification is
useful to guess a good labelling of the given graph. The corresponding
MSO-sentence is known from the knowledge of the bounds of pp-cographs.
However, the class of probe cographs is FO-definable by Proposition 2.4(2),
but we do not know the corresponding sentence as the bounds of probe cographs
are not completely known. $\ \square$

\bigskip

\textbf{Theorem 5.4\ :} There is an algorithm that can compute the finitely
many bounds of the class of probe cographs. An upper-bound to their sizes\ is computable.

\textbf{Proof sketch : }By Proposition 5.3, we can construct effectively an
MSO-sentence $\xi$ that defines the class $\mathcal{B}$\ of bounds of probe
cographs among finite graphs.\ By Theorem 6.35 of \cite{CouEng}\ or an
algebraic version of it in terms of recognizable sets (Corollary 5.59), one
can build a finite automaton $\mathcal{A}$\ that recognizes the set of normal
terms of width at most 8 that define the graphs in $\mathcal{B}$.\ Then
$L(\mathcal{A})$ is finite as we know that $\mathcal{B}_{\simeq}$
is.\ However, several terms in $L(\mathcal{A})$ may define isomorphic
graphs.\ As $L(\mathcal{A})$ is finite, one can list its elements and thus the
graphs it defines after removing isomorphic duplicates.

The MSO-sentence $\xi$\ can be replaced by $\xi$ $\wedge\delta$ where $\delta$
is the MSO-sentence expressing that a graph is connected and has a diameter at
most 4.\ We obtain in this way a more restrictive set $L(\mathcal{A})$ without
missing any graph in $\mathcal{B}$ except $P_{6}$, but we know it.

\emph{Pumping lemmas} are classical tools of language theory by which one can
bound the sizes of the terms of a finite recognizable set defined by a given
finite automaton without listing them. However the obtained bound would be
ridiculous huge.\ \ $\square$

\bigskip

This decision procedure is actually intractable, because of the complexity of
the sentence $\xi$ and the size of the corresponding automaton, that needs to
handle clique-width terms with 8\ labels.\ 

\bigskip

\emph{Some bounds of probe cographs}

By a \emph{bound}, we mean a bound of probe cographs, hence, a graph in
$\mathcal{B}$.

We denote by $\overline{G}$ the edge-complement of a graph $G$. By
\emph{substituting an edge} (\emph{i.e.}, the graph $K_{2}$) to a vertex $a$
of a graph $G$, we obtain the graph denoted by $G[a\longleftarrow K_{2}%
]$.\ Its vertex $a$ is replaced by the edge $a_{1}-a_{2}$ and any edge $a-x$
of $G$ is replaced by the two edges $a_{1}-x$ and\ $a_{2}-x$.\ Note that
\ $G[a\longleftarrow K_{2}]-a_{1}$ is isomorphic to $G$. We have
$G[a\longleftarrow K_{2}][b\longleftarrow K_{2}]=G[b\longleftarrow
K_{2}][a\longleftarrow K_{2}]$ if $b\neq a$.\ It is the result of the
\emph{simultaneous substitution} of $K_{2}$ for $a$ and $b$, denoted by
$G[a\longleftarrow K_{2},b\longleftarrow K_{2}].$

\bigskip

\textbf{Proposition 5.5\ }: The following graphs, all of clique-width 3, are
in $\mathcal{B}$ :

(1) The standard graphs $C_{5},P_{6}$ and $C_{6}.$

(2) The graphs $\overline{C_{6}}$, $D$ and $\overline{D}$ derived from $C_{6}%
$, see Figure 3.

(3) Four graphs obtained by substituting an edge to one or two vertices of a
path $P_{4}$ or $P_{5}$.\ See Figure 4.

(4) Two graphs obtained from the \emph{house} $H$\ by substituting edges as in
(3).\ See Figure 5.

(5) Two graphs obtained as in (4) from two p-cographs with 5 vertices. See
Figure 6.

\textbf{Proof}: The proofs are based on the following observations
(cf.\ Example 2.5)~:

\begin{quote}
the only good labellings of $P_{4}$ are 1212, 2121 and 1221,

the only good labelling of $P_{5}$ is 12121,

the two good labellings of the "house" $H$\ with vertices $\{a,b,c,d,e\}$ and
top vertex $c$, cf.\ Figure 5, label by 1\ either $c$ and $d,$ or $c$ and $b$;
the other vertices are labelled by 2,

every good labelling of a graph is good for its induced subgraphs.
\end{quote}

(1) See Examples 2.5 for $C_{5}$ and $P_{6}$. Let $C_{6}\ =a-b-c-d-e-f-a$%
.\ Assume for a contradiction that it has a good labelling.\ By removing $a$,
we get $P_{5}$, hence $b$ and $f$ must be labelled by 1, hence $a$ must be
labelled by 2. Similarly, $b$ must be labelled by 2. As $P_{5}$ is a
p-cograph, $C_{6}$ is a bound. Hence, $C_{5},P_{6}$ and $C_{6}$ are bounds.

(2) The graph $\overline{C_{6}}$ is shown in Figure 3.\ By removing $f$, we
get the "house" $H$ with top vertex $c$ that should be labelled by 1, so that
either $b$ or $d$ should be labelled by 1. Hence, $f$ must be labelled by 2,
but it is the top vertex of the house $\overline{C_{6}}-c.\ $Hence,
$\overline{C_{6}}$ is not a p-cograph.\ However, it is a bound.

The graph $D$\ is obtained from $C_{6}\ =a-b-c-d-e-f-a$\ by adding the edge
$c-f$ .\ By removing $f$, we see that $a,c$ and $e$ should be labelled by 1,
and $f$ by 2. If we remove $c$, we get a path $P_{5}$ of type 21212 which is
not good.\ 

From the graph $\overline{D}$\ shown in Figure 3, we get two "houses" by
removing either $c$ or $f$.\ Both vertices should be labelled by 1, so that
none of the others can be labelled by 1.\ 

\bigskip%
\begin{figure}
[ptb]
\begin{center}
\includegraphics[
height=2.4491in,
width=1.8628in
]%
{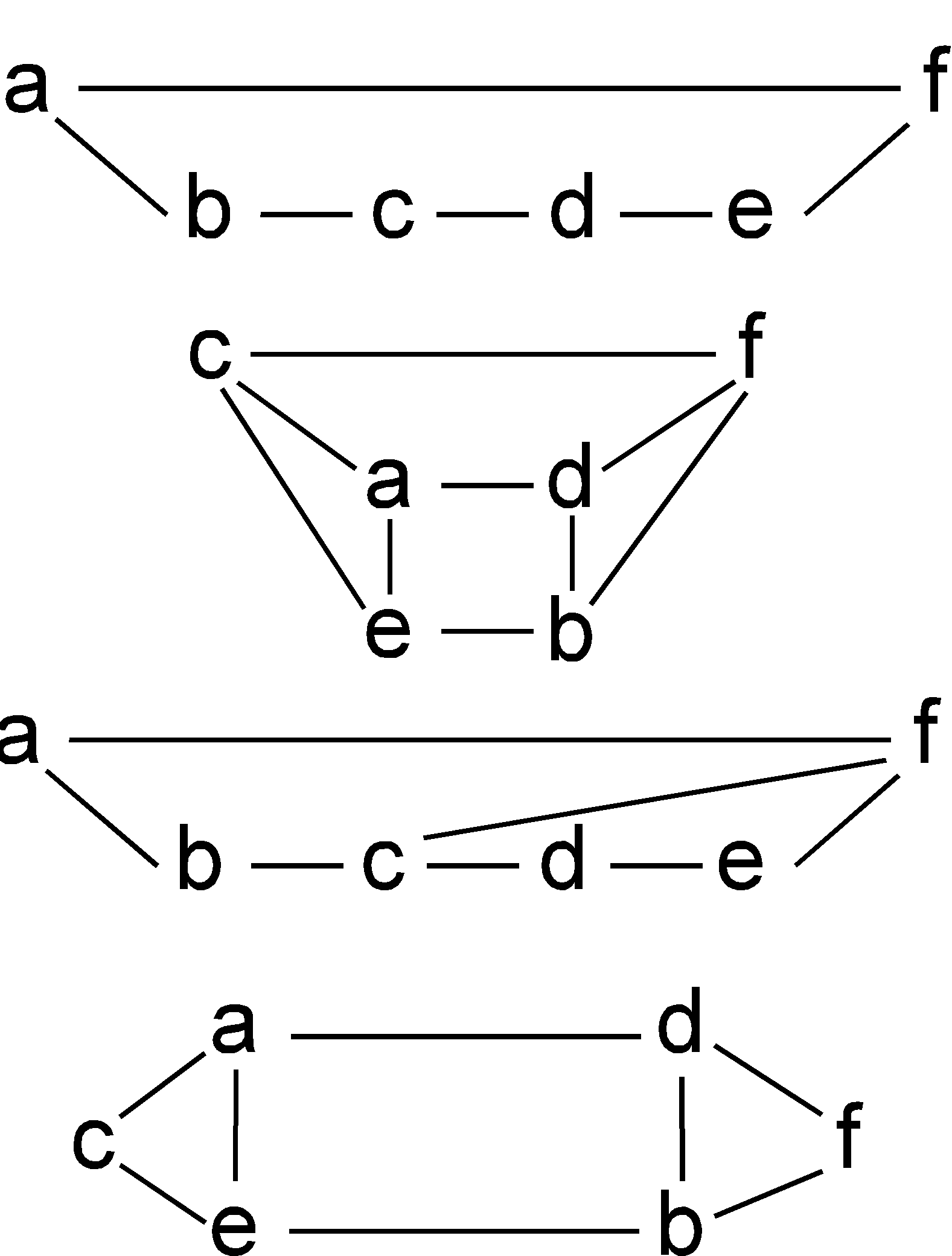}%
\caption{The bounds $C_{6}$, $\overline{C_{6}}$, $D$ and $\overline{D}$ of
Proposition 5.5(1,2).}%
\end{center}
\end{figure}

(3) The path $P_{5}=a-b-c-d-e$ has a unique good labelling of type 12121.\ If
we substitute $K_{2}$ for any of $a,c$ or $e$, we obtain a bound as the two
vertices of the substituted edge cannot be both labelled by 1. We obtain only
two non-isomorphic bounds, shown in the top part of Figure 4, because
substituting an edge to $a$ and $e$ give isomorphic graphs.

The path $P_{4}=a-b-c-d$ has three good labellings of types 1212, 2121 and
1221. For each good labelling, at least one vertex in $\{a,b\},$ in $\{a,d\}$
and in $\{c,d\}$ must be labelled by 1.\ It follows that $P_{4}%
[a\longleftarrow K_{2},b\longleftarrow K_{2}]$ and $P_{4}[a\longleftarrow
K_{2},d\longleftarrow K_{2}]$ shown in the bottom part of Figure 4 are bounds
as one checks easily.\ So is $P_{4}[c\longleftarrow K_{2},d\longleftarrow
K_{2}]$ isomorphic to the first one.\ We obtain two non-isomorphic bounds.%

\begin{figure}
[ptb]
\begin{center}
\includegraphics[
height=2.3263in,
width=1.5022in
]%
{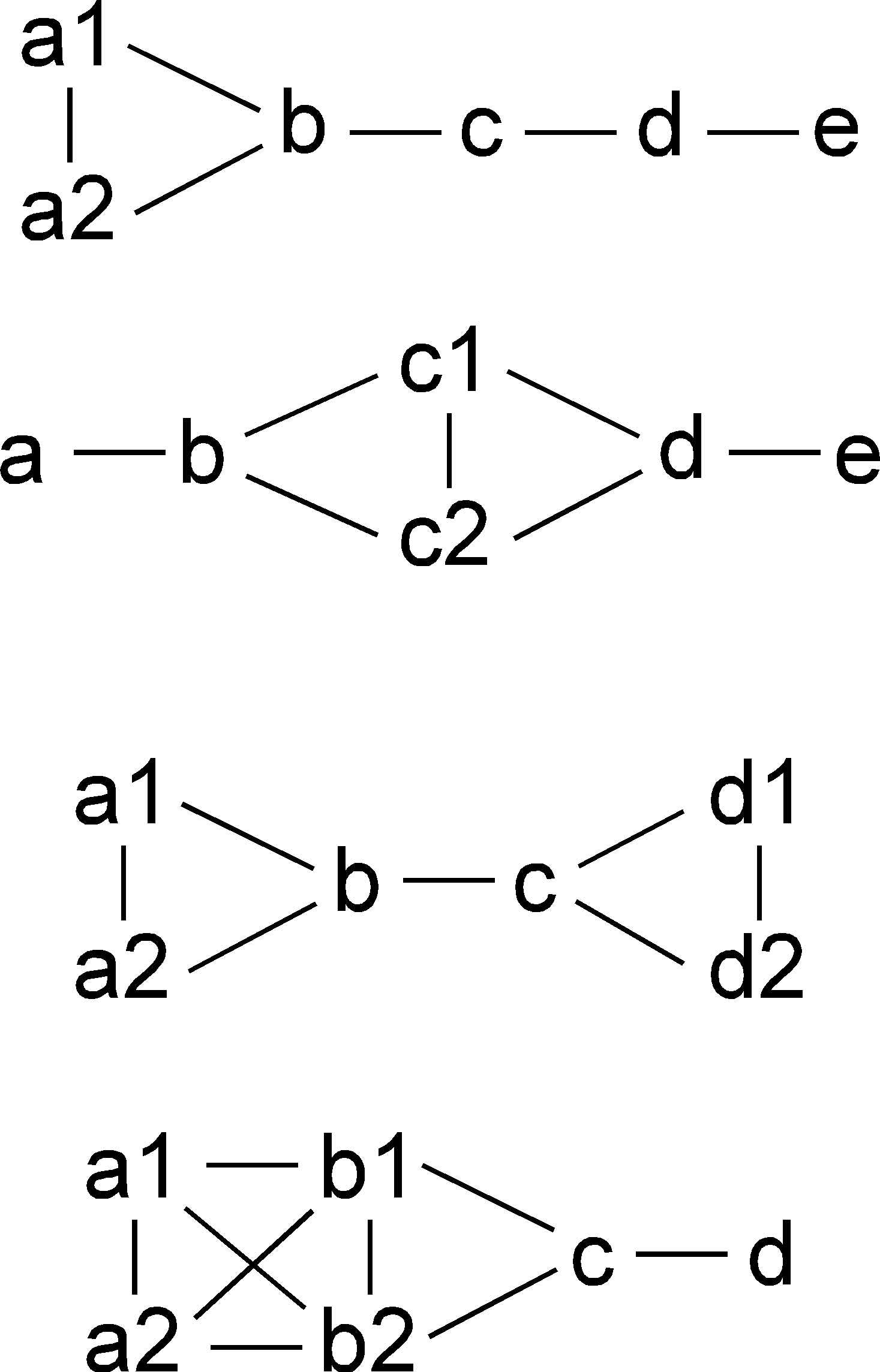}%
\caption{The bounds of Proposition 5.5(3).}%
\end{center}
\end{figure}

(4) Every good labelling of the "house" $H$ shown to the left of Figure 5,
must label $c$ by 1\ and, either $b$ or $d,$ by 1. We obtain the two bounds
$H[c\longleftarrow K_{2}]$ and $H[b\longleftarrow K_{2},d\longleftarrow
K_{2}]$ shown in Figure $4$.%

\begin{figure}
[ptb]
\begin{center}
\includegraphics[
height=1.2332in,
width=3.141in
]%
{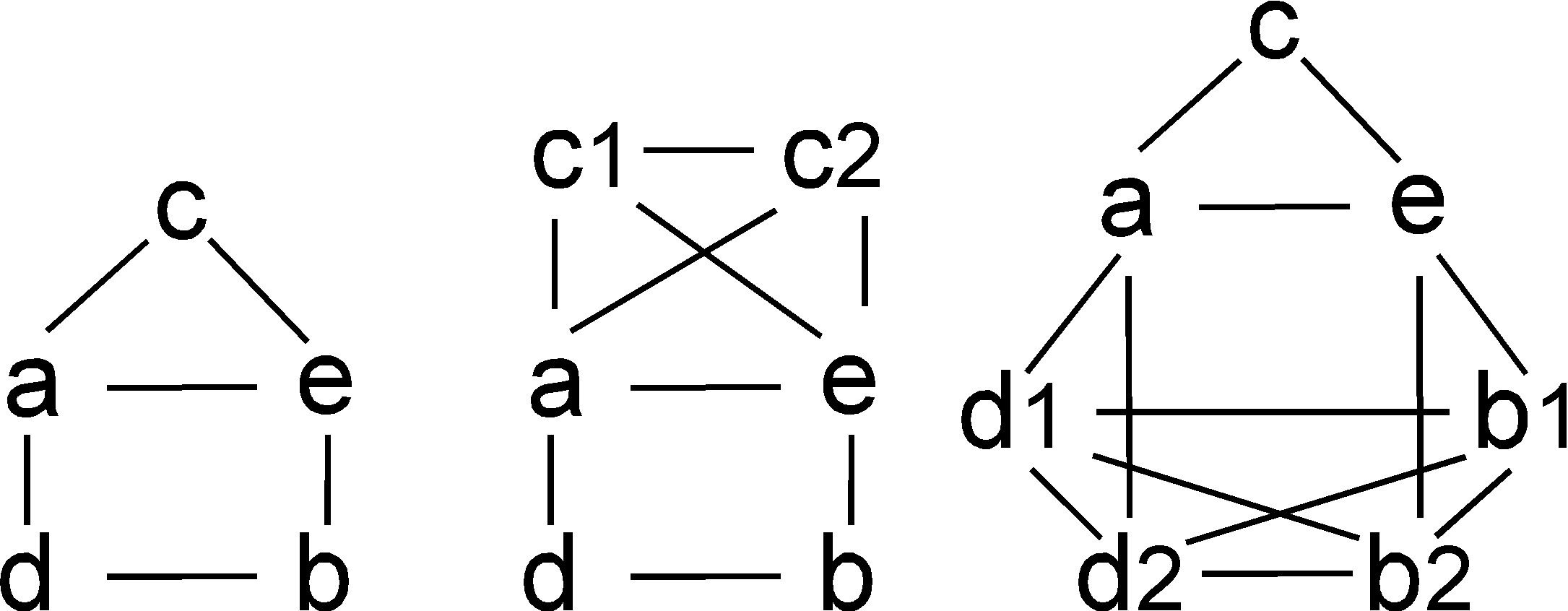}%
\caption{The "house" to the left (it is a p-cograph), and the two bounds of
Proposition 5.5(4).}%
\end{center}
\end{figure}

(5) Among the connected graphs with 5 vertices that are not cographs we have
are $C_{5}$ and $H$. Figure 6 shows two others, $H_{1}$ and $H_{2}$.

Every good labelling of $H_{1}$, labels $b$ or $d$ (or both) by 1.\ Hence
$H_{1}[b\longleftarrow K_{2},d\longleftarrow K_{2}]$ is not a p-cograph. It is
a bound because $H_{1}[b\longleftarrow K_{2}]$ and $H_{1}[d\longleftarrow
K_{2}]$ are p-cographs.

Similarly, we get from $H_{2}$, the bound $H_{2}[c\longleftarrow
K_{2},d\longleftarrow K_{2}]$.

\bigskip

The graphs $C_{5},P_{6}$ and $C_{6}$\ are known to have clique-width 3. That
$H$, $\overline{C_{6}}$, $D$ and $\overline{D}$ have clique-width 3 can be
checked with \cite{TRAG} or proved directly.\ All other bounds of (3),(4) and
(5) are obtained by substituting $K_{2}$\ to vertices of graphs of
clique-width 3.\ Hence, they have clique-width 3.\ \ \ \ $\square$

We do not know any graph of clique-width 4 or more that is a bound of probe
cographs.\ Hence, for now, we are far from the upper-bound 8 of Proposition 5.3.

%

\begin{figure}
[ptb]
\begin{center}
\includegraphics[
height=1.3569in,
width=2.2978in
]%
{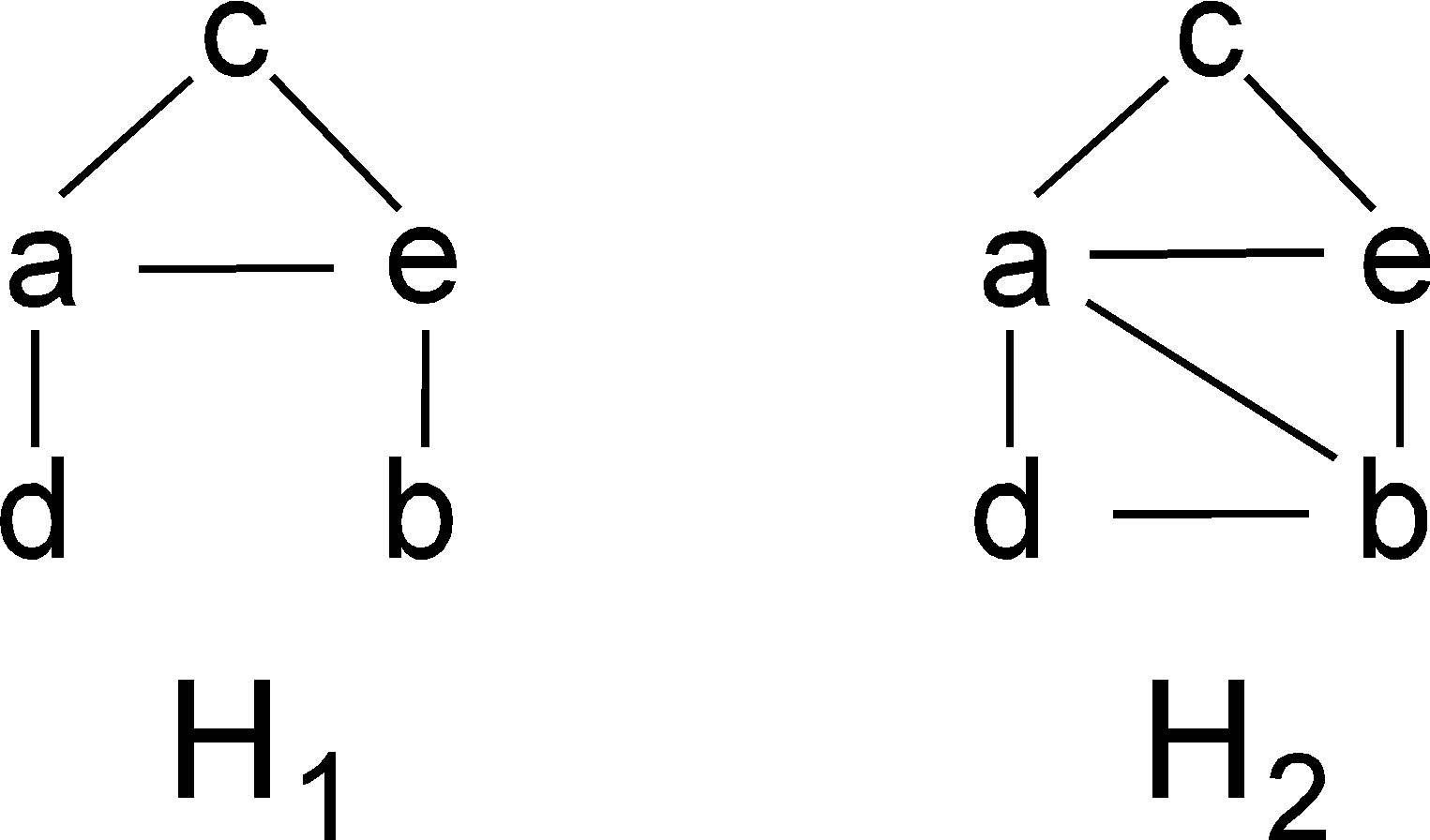}%
\caption{Two p-cographs used in Proposition 5.5(5)}%
\end{center}
\end{figure}

\section{Open problems}

\textbf{Problem 6.1}\ : Determine the set of bounds of probe cographs. What
can be said about them in addition to what is stated in Proposition 5.3 ?

\bigskip

Monadic second-order\ logic does not help for effective computations because
of the huge sizes of the automata constructed from MSO-sentences.

\bigskip

\textbf{Problem\ 6.2 }: Does there exist a \emph{monadic second-order
transduction}, cf.\ Chapter 7 of \cite{CouEng}, that transforms a finite
ternary structure assumed to be in \textbf{IBO}\ into a finite marked tree
defining it?

\bigskip

In \cite{CouLMCS2020}, we studied four classes of betweenness structures (cf.
Definition 4.1) :\ \textbf{QT},\textbf{ IBQT}, \textbf{BO}\ and \textbf{IBO}.
Each betweenness structure $S=(N,B)$ is defined from a labelled O-tree, say
$T=(M,\leq,N_{\oplus},N_{\otimes}).$ This description covers all cases,
although labels are useless in some cases.\ 

The question is whether some witnessing O-tree $T$\ can be defined by
MSO-formulas in the given structure $S$, in technical words, by an
MSO-transduction. An FO-transduction exists for \textbf{QT} and
MSO-transductions exist for \textbf{IBQT} and \textbf{BO} \cite{CouLMCS2020}.

Theorem 4.2(1)\ establishes that the class \textbf{IBO}\ is MSO$_{fin}%
$-definable, without building an associated MSO-transduction. We recall that
an MSO-transduction transforms a structure with $n$ elements into one with at
most $kn$ elements for some fixed $k$. As a finite structure in \textbf{IBO}%
\ having $n$ elements can be defined from a marked tree with at most $2n-1$
nodes, it is not hopeless to look for such a transduction.\ An intermediate
result seems necessary: to find an MSO-transduction that constructs, from a
pp-cograph, a term that defines it.\ Such a transduction exists for countable
cographs given with an auxiliary linear order \cite{CouX,CouDel}.\ Such an
auxiliary order would be useful, even perhaps necessary.

\bigskip

\textbf{Problem 6.3}\ : Determine the set of bounds of \textbf{IBO}.

\bigskip

We have presently no result similar to Theorem 5.4 because ternary structures
do not share certain good properties of graphs, as we now explain.

First we observe that, since we have an effective MSO characterization of the
finite structures in \textbf{IBO} by Theorem 4.2(1), we have one of the set of
their bounds because the proof for graphs (Proposition 5.3) extends to
relational structures.\ 

The proof of Theorem 5.4 uses the fact that the bounds of probe cographs have
clique-width $\leq8$ (even if this upper-bound is overestimated).\ We miss a
corresponding fact for $Bnd$(\textbf{IBO}).\ First because there is no really
convenient notion of clique-width for ternary structures.\ However, we can
replace the property "the graphs of $\mathcal{C}$ have clique-width at most
$k$" by "the structures of $\mathcal{C}$ are all in $\tau$(\textbf{Trees}) for
some MSO-transduction $\tau$" where \textbf{Trees }is the class of finite
rooted trees, and say that $\mathcal{C}$ is \emph{tree-definable}. Bounded
clique-width is equivalent for a class of graphs to tree-definability
(\cite{CouEng}, Chapter 7). Furthermore, the computability results for classes
of graphs of bounded clique-width hold for tree-definable classes of structures.

For a set $\mathcal{C}$ of structures of a fixed signature, we let
$\mathcal{C}^{+}$\ be the set of structures $S$ with domain $N$ such that
$S[N-x]$ is in $\mathcal{C}$\ for some $x$ in $N$.\ If $\mathcal{C}$ is a set
of graphs of clique-width at most $k$, then the graphs in $\mathcal{C}^{+}%
$\ have clique-width at most $2k$ (see \cite{Gur}), which is used to obtain
the upper-bound 8\ to the clique-width of the bounds of probe cographs in
Proposition 5.3. However, this fact does not extend to tree-definable classes
of structures as we prove below. Hence, we cannot extend Theorem 5.4\ to the
computation of $Bnd$(\textbf{IBO}).\ 

\bigskip

\textbf{Proposition 6.4} : There is a tree-definable set of ternary structures
$\mathcal{C}$ $\subseteq$ \textbf{IBO} such that $\mathcal{C}^{+}$\ is not tree-definable.

\textbf{Proof} : We will use results from \cite{CouEng}.\ Let $S_{G}%
=([n],B_{G})$ where $B_{G}$\ consists of the triples $(1,i,j)$ and $(j,i,1)$
for the edges $i-j$ of some graph $G=(\{2,3,...,n\},E)$ and $i<j$. Then
$S_{G}$ is a ternary structure that satisfies Properties A1-A6.\ It is in
$\mathcal{C}^{+}$ where $\mathcal{C}$\ is the set of trivial ternary
structures $(\{2,3,...,n\},\emptyset)$, obviously in \textbf{IBO}. There is an
MSO-transduction $\theta$ that transforms each structure $S_{G}$ into $G$\ :
it deletes 1 and replaces the triples $(1,i,j)$ and $(j,i,1)$ in $B_{G}$\ by
$(i,j)$ and $(j,i),$ thus defining $E$.

It is clear that $\mathcal{C}$\ is tree-definable.\ If $\mathcal{C}^{+}$ would
be, that is, if $\mathcal{C}^{+}$ $\subseteq$ $\tau$(\textbf{Trees}) for some
MSO transduction $\tau,$\ then the MSO transduction $\theta\circ\tau$ would
produce all finite graphs from \textbf{Trees} (up to isomorphism), hence, all
graphs would have clique-width bounded by a fixed value (\cite{CouEng},
Chapter 7), which is false. $\square$

\end{document}